\documentclass[12pt,numbers]{article}

\usepackage{natbib}
\usepackage{verbatim}
\usepackage{amsmath}
\usepackage{amsthm}
\usepackage{amsfonts}
\usepackage{amssymb}
\usepackage[all]{xy}     
\usepackage{color}
\usepackage[title]{appendix}

\usepackage[colorlinks=false,linktocpage]{hyperref}

\usepackage{feynmp}
\newlength{\xtrawidth}
\newlength{\xtraheight}
\def\clap#1{\hbox to 0pt{\hss#1\hss}}

{\end{list}}
{\everymath{\displaystyle\everymath{}}\array}%
{\endarray}

\newcommand{\eqdef}{%
  \mathrel{\lower.1mm
    \hbox{$\stackrel{\lower.424ex\hbox{\scriptsize def}}{=}$}}
}
\DeclareMathOperator{\Tr}{Tr}

\DeclareMathOperator{\ad}{ad}

\DeclareMathOperator{\rank}{rank}

\DeclareMathOperator{\Span}{span}

\newcommand{\dslash}{\mbox{$\partial$ \kern-.92em  \big /}}
\newcommand{\diff}{\mathrm{d}}
\newcommand{\Z}{\mathbb{Z}}

\newcommand{\C}{\mathbb{C}}

\newcommand{\CP}[1]{\mathbb{P}^{#1}}
\newcommand{\IP}[1]{\CP{#1}}

\newcommand{\ZZZ}{{\ensuremath{\Z_3\times\Z_3}}}
\newcommand{\Rep}[1]{\ensuremath{\mathbf{#1}}}
\newcommand{\barRep}[1]{\ensuremath{\overline{\Rep{#1}}}}

\newcommand{\dP}[1]{d\mathbb{P}_{#1}}

\newcommand{\B}[1]{\ensuremath{B_{#1}}}
\newcommand{\Xt}{{\ensuremath{\widetilde{X}}}}
\newcommand{\V}[1]{\ensuremath{{V}_{#1}}}
\newcommand{\Vt}{\ensuremath{\widetilde{V}}}
\newcommand{\W}[1]{{\ensuremath{{W}_{#1}}}}
\newcommand{\Osheaf}{\ensuremath{\mathcal{O}}}

\newcommand{\oB}[1]{\ensuremath{\Osheaf_{\B{#1}}}}
\newcommand{\oXt}{\ensuremath{\Osheaf_{\Xt}}}
\newcommand{\p}[1]{{\ensuremath{\pi_{#1}}}}

\newcommand{\Fsheaf}{\ensuremath{\mathcal{F}}}

\definecolor{grey}{gray}{0.4}
\newcommand{\Hpq}[3]{\big({#2},{#3}\textcolor{grey}{\big|{#1}}\big)}
\newcommand{\Hst}[4]{\big[{#3},{#4}\textcolor{grey}{\big|{#2},{#1}}\big]}

\newcommand{\cO}{\Osheaf}
\newcommand{\cF}{\Fsheaf}
\newcommand{\cFt}{\widehat{\Fsheaf}}

\newcommand{\atv}{\wedge^2 \Vt}
\newcommand{\tv}{\Vt}
\newcommand{\tx}{\Xt}
\def\z3z3{\ZZZ}

\begin{document}

\begin{titlepage}
    \vspace*{\stretch{0}}
  \begin{center}
     \Huge 
     Two Higgs Pair Heterotic Vacua\\ and Flavor-Changing Neutral Currents
  \end{center}
  \vspace*{\stretch{0.5}}
  \begin{center}
    \begin{minipage}{\textwidth}
      \begin{center}
        \Large
        Michael Ambroso,         
        Volker Braun, 
        Burt A.~Ovrut
      \end{center}
    \end{minipage}
  \end{center}
  \begin{center}
    \begin{minipage}{\textwidth}
      \begin{center}
         Department of Physics
        \\
        University of Pennsylvania
        \\        
        Philadelphia, PA 19104--6395, USA
      \end{center}
      
    \end{minipage}
  \end{center}
  \vspace*{4mm}
  \begin{abstract}
    \normalsize 
We present a vacuum of heterotic $M$-theory whose observable sector has the MSSM spectrum 
with the addition of one extra pair of Higgs-Higgs conjugate superfields. The quarks/leptons have a realistic mass hierarchy with a naturally light first family. The double elliptic structure of the Calabi-Yau compactification threefold leads to two ``stringy'' selection rules. These classically disallow Yukawa couplings to the second Higgs pair and, hence, Higgs mediated flavor-changing neutral currents. Such currents are induced in higher-dimensional interactions, but are naturally suppressed. We 
show that our results fit comfortably below the observed upper bounds on neutral flavor-changing processes.
  \end{abstract}
  \vspace*{\stretch{5}}
  \begin{minipage}{\textwidth}
    \underline{\hspace{5cm}}
    \centering
    \\
    Email: 
    \texttt{mambroso@sas.upenn.edu},
    \texttt{vbraun@physics.upenn.edu},
    \texttt{ovrut@elcapitan.hep.upenn.edu}.
  \end{minipage}
\end{titlepage}

\section{Introduction}

Heterotic $M$-theory~\cite{hw1,het1,het2} offers a venue for finding phenomenolgically realistic vacua of the heterotic string~\cite{hs1}.
Although several different approaches are possible, see for example~\cite{oa1,oa2,oa3}, the construction of non-standard embedded 
holomorphic vector bundles on elliptically fibered Calabi-Yau 
threefolds~\cite{fmw1,bund1,math1,math2} has proven to be particularly fruitful. Within this context, one can explicitly compute the zero-mode spectrum using sheaf cohomology~\cite{spec1}.
Chiral quark/lepton three family vacua with natural doublet-triplet Higgs splitting~\cite{spec2} and no exotic quantum number fields are easily achieved. Generically, these also contain both vector-like pairs of matter fields as well as several vector-like pairs of Higgs superfields. However, vacua can be constructed with no vector-like pairs of matter, of which a substantial subset have at most two Higgs pairs~\cite{hsm1,hsm1a}. Furthermore, there are a small number of such vacuum states with only one Higgs pair~\cite{hsm2,hsm2a}; that is, with exactly the spectrum of the MSSM. We have called these ``Heterotic Standard Models''.

Finding heterotic vacua with a spectrum either exactly, or close to, the MSSM is just the beginning of the physical analysis. It is crucial that the perturbative cubic couplings of the zero-mode fields, that is, the coupling of Higgs-Higgs  conjugate fields to moduli and the cubic Yukawa terms, lead to realistic 
$\mu$-terms and fermion mass matrices respectively. The texture of these couplings can be determined by evaluating the cubic cohomology products using Leray spectral sequences~\cite{lss,lssa,lssb}. It was shown that many Heterotic Standard Models have naturally suppressed $\mu$-terms and a realistic hierarchy of physical masses. Furthermore, one must compute the non-perturbative string corrections to the moduli superpotential so as to stabilize the vacuum. This has been carried out in a series of papers~\cite{np1,np2}. Having fixed the geometric and vector bundle moduli, it is possible, using both mathematical and numerical methods~\cite{don1}, to compute the explicit metrics on the Calabi-Yau threefolds~\cite{cy1}, the eigenspectra of bundle valued Laplacians on these spaces~\cite{lap1} and, using these results, the explicit $\mu$-term coefficients and Yukawa couplings. This latter calculation is in progress.

It is of interest to note that it is substantially easier to find Heterotic Standard Models with two Higgs-Higgs conjugate pairs than such vacua with only the single Higgs pair of the MSSM. The reason is rooted in the associated algebraic geometry. At the end of the day, it is less of a constraint to impose that there be two Higgs pairs and, hence, we find many more such vacua. It is of relevance, therefore, to explore the physical properties of these two Higgs pair Heterotic Standard Models and to exhibit vacua with reasonable physical characteristics, such as a realistic fermion mass matrix. This will be carried out in this paper. Using an extension of the methods presented in~\cite{spec1,hsm1,hsm1a,hsm2,hsm2a}, we construct a class of heterotic $M$-theory vacua whose observable sector has the spectrum of the MSSM with the addition of a second Higgs pair. There are no other vector-like pairs of fields or fields with exotic quantum numbers. This two Higgs pair Heterotic Standard Model is shown to have an acceptable hierarchical mass spectrum with a very light first family. 

The addition of the second Higgs pair poses the serious problem of potentially generating large Higgs mediated flavor-changing neutral currents. Since a number of such processes have strict experimental upper bounds, this concern must be addressed. We do that in this paper. First, we show that the ``stringy'',  so-called $(p,q)$ and $[s,t]$, selection rules that arise from two Leray spectral 
sequences~\cite{lssa} disallow all matter couplings to the second Higgs pair classically. That is, 
all classical flavor-changing neutral currents naturally {\it vanish}. Such interactions can arise from the coupling of the zero-mode fields to the massive Kaluza-Klein tower of states, but these neutral current interactions are of higher order in the fields and, hence, are {\it naturally suppressed}. Using a non-supersymmetric two Higgs doublet ``toy'' model, which none-the-less captures the relevant features of the two Higgs pair supersymmetric vacuum, we show that the Higgs mediated flavor-changing neutral currents generated by the second Higgs-Higgs conjugate pair sit comfortably below the present experimental upper bounds. We briefly discuss a possible region of parameter space where the two Higgs pair vacua could induce flavor-changing phenomena approaching the experimental upper bound of some processes.

Specifically, we do the following. In Section 2, we present the explicit elliptically fibered Calabi-Yau
threefold and $SU(4)$ holomorphic vector bundle of our two Higgs pair vacua. Using techniques introduced in~\cite{spec1,hsm1,hsm1a,hsm2, hsm2a}, the spectrum is shown to be precisely that of the MSSM with the addition of a second Higgs-Higgs conjugate pair. We also compute the number of geometric and vector bundle moduli; 
$h^{1,1}(X)=h^{2,1}(X)=3$ and 13 respectively. The texture of the cubic Yukawa terms in the superpotential is calculated in Section 3. These terms are shown to arise as the cubic product of the sheaf cohomology groups associated to matter and Higgs-Higgs conjugate superfields. The internal properties of these cohomologies under the $(p,q)$ and $[s,t]$ ``stringy'' symmetries induced by the two Leray sequences are tabulated and shown to lead to explicit selection rules for these couplings. The associated texture of the quark/lepton mass matrix is computed explicitly and found to naturally have one light and two heavy families. Importantly, we show that the stringy symmetries allow the coupling of left and right chiral matter to the first Higgs pair but disallow a cubic coupling of matter to the second Higgs-Higgs conjugate superfields. Thus, classically, these two Higgs pair Heterotic Standard Models have no flavor-changing neutral currents.
In Section 4, a similar calculation is carried out for the cubic terms in the superpotential involving a single vector bundle modulus with the Higgs-Higgs conjugate pairs. The $(p,q)$ and $[s,t]$ symmetries of the associated sheaf cohomologies again induce a texture on these couplings, allowing only 9 of the 13 vector bundle moduli to form such couplings and restricting the Higgs content as well. This has important consequences for the magnitude of the Higgs induced flavor-changing neutral currents. 

A discussion of the superpotential, including a heavy Kaluza-Klein superfield and its cubic coupling to two zero-mode fields, is given in Section 5. It is shown that tree level supergraphs involving the exchange of a Kalaza-Klein superfield can generate the coupling of quark/lepton chiral matter to the second Higgs-Higgs conjugate pair, but only at dimension 4 in the superpotential. Hence, there is a natural suppression by a factor of $1/M_c$, where $M_{c}$ is the compactification scale. Similarly, such supergraphs generate suppressed dimension 4 terms in the superpotential coupling all 13 vector bundle moduli to all Higgs pairs. By requiring that these vacua have the correct scale of electroweak symmetry breaking, one can put an upper bound on the size of the vector bundle moduli vacuum expectation values and, hence, on the magnitude of the Yukawa couplings to the second Higgs-Higgs conjugate pair. Finally, in Section 6, we represent the physics of our two Higgs pair vacua in terms of a simplified model. This is essentially the non-supersymmetric standard model with the addition of a second Higgs doublet and a real scalar field representing the 4 vector bundle moduli disallowed from forming cubic couplings. The fact that chiral matter is prevented classically from coupling to the second Higgs pair is enforced in the toy model by a ${\mathbb{Z}}_{2}$ symmetry~\cite{weinberg}. The scalar vacuum state closest to that of the standard model is found and the associated Higgs and fermion masses and eigenstates computed. Using these, we compute the interaction Langrangian for the Higgs mediated flavor-changing neutral currents, constraining the coefficients of these interactions to be those determined in Section 5 in the supersymmetric string vacua. These interactions are compared with the experimental upper bounds in several $\Delta F=2$ neutral meson processes~\cite{atwood1,atwood2} and found to be generically well below these bounds. However, by choosing  certain parameters to be of order unity, and for a sufficiently light neutral Higgs scalar, the flavor-changing neutral current contributions to some meson processes can approach the upper bounds.

\section{The Two Higgs Pair Vacuum}

We now specify, in more detail, the properties of these 
vacua with two Higgs-Higgs conjugate pairs
and indicate how they are determined. The requisite Calabi-Yau
threefold, $X$, is constructed as follows~\cite{hsm1}.  Let $\Xt$ be a
simply connected Calabi-Yau threefold which is an elliptic fibration
over a rational elliptic surface, $\dP9$. It was shown
in~\cite{rein1} that $\Xt$ factors into the fibered product $\Xt=B_{1}
\times_{\mathbb{P}^1} B_{2}$, where $B_{1}$ and $B_{2}$ are both
$\dP9$ surfaces. Furthermore, $\Xt$ is elliptically fibered with
respect to each projection map $\pi_{i}:\Xt \rightarrow B_{i}$,
$i=1,2$. In a restricted region of their moduli space, such manifolds
can be shown to admit a $\ZZZ$ group action which is fixed-point free.
It follows that
\begin{equation}
  X=\frac{\Xt}{\ZZZ}
  \label{1}
\end{equation}
is a smooth Calabi-Yau threefold that is torus-fibered over a singular
$d\mathbb{P}_9$ and has non-trivial fundamental group
\begin{equation}
  \pi_{1}(X)=\ZZZ
  \,,
  \label{2}
\end{equation}
as desired. It was shown in~\cite{hsm1} that $X$ has 
\begin{equation}
  h^{1,1}(X)=3 \,, \quad 
  h^{2,1}(X)=3
  \label{3}
\end{equation}
Kahler and complex structure moduli respectively; that is, a total of $6$
geometric moduli.

We now construct a holomorphic vector bundle, $\V{}$, on $X$
with structure group
\begin{equation}
  G= SU(4)  
  \label{4}
\end{equation}
contained in the $E_8$ of the observable sector. For this bundle 
to admit a gauge connection satisfying the hermitian
Yang-Mills equations, it must be slope-stable. The connection 
spontaneously breaks the observable sector $E_8$ gauge symmetry to
\begin{equation}
  E_8 \longrightarrow Spin(10)
  \,,
  \label{5}
\end{equation}
as desired. We produce $\V{}$ by building stable, holomorphic vector bundles
$\Vt$ with structure group $SU(4)$ over $\Xt$ that are equivariant
under the action of $\ZZZ$. This is accomplished by generalizing the
method of ``bundle extensions'' introduced in~\cite{extension}. The
bundle $\V{}$ is then given as
\begin{equation}
  V=\frac\Vt{\ZZZ}
  \,.
  \label{6}
\end{equation}

Realistic particle physics phenomenology imposes additional
constraints on $\Vt$. Recall that with
respect to $SU(4) \times Spin(10)$ the adjoint representation of $E_8$
decomposes as
\begin{equation}
  \Rep{248}=
  \big( \Rep{1},\Rep{45} \big) \oplus  
  \big( \Rep{4},\Rep{16} \big) \oplus 
  \big( \barRep{4}, \barRep{16} \big) \oplus 
  \big( \Rep{6},\Rep{10} \big) \oplus
  \big( \Rep{15},\Rep{1} \big)  
  \,.
  \label{8}
\end{equation}
The low-energy spectrum arising from compactifying on $\tx$ with vector
bundle $\tv$ is determined from ~\cite{spec1} 
\begin{multline}
	\ker\big(\dslash_{\Vt}\big) = \left( 
    H^0\big(\tx, \cO_{\tx}\big) \otimes \Rep{45} \right)
  \oplus \left( H^1\big(\tx, \tv^\ast \big) \otimes
    \barRep{16} \right) \\ 
  \oplus
  \left( H^1\big(\tx, \tv\big) \otimes
    \Rep{16} \right) \oplus 
  \left( H^1\big(\tx, \atv \big) \otimes \Rep{10} \right)
  \oplus 
  \left( H^1\big(\tx, \ad(\tv) \big) \otimes
    \Rep{1} \right),
\end{multline}
where $\dslash_{\tv}$ is the Dirac operator on $\tx$ twisted
by $\tv$.  The multiplicity of each representation $R$ of $Spin(10)$
is the dimension of the associated cohomology space.

The number of $\Rep{45}$ multiplets is given by
\begin{equation}
  h^{0}\left(\Xt, \oXt \right)=1.
  \label{yes1}
\end{equation}
Hence, there are $Spin(10)$ gauge fields in the low-energy theory, but
no adjoint Higgs multiplets. The chiral families of quarks/leptons
will descend from the excess of $\Rep{16}$ over $\barRep{16}$
representations. To ensure that there are three generations of quarks
and leptons after quotienting out $\ZZZ$, one must require that
\begin{equation}
  n_{\barRep{16}} -
  n_{\Rep{16}}
  =
  \frac{1}{2} c_{3}\big( \Vt \big)
  =
  -3 \cdot \big| \ZZZ \big| 
  = 
  -27
  \,,
  \label{7}
\end{equation}
where $n_{\barRep{16}}$, $n_{\Rep{16}}$ are the numbers of
$\barRep{16}$ and $\Rep{16}$ multiplets respectively, and
$c_{3}(\Vt)$ is the third Chern class of $\Vt$.  
Furthermore, if we demand that there be no
vector-like matter fields arising from $\barRep{16}$-$\Rep{16}$ pairs,
$\Vt$ must be constrained so that
\begin{equation}
  h^1\left( \Xt, \Vt^\ast \right)=0
  \,.
  \label{9}
\end{equation}
Similarly, the number of $\Rep{10}$ zero modes is
$h^1\big(\Xt,{\wedge}^{2}\Vt\big)$. However, since the Higgs fields
arise from the decomposition of the $\Rep{10}$, one must not set the
associated cohomology to zero.

In~\cite{hsm2}, it was shown that the \emph{minimal}, non-vanishing number of $\Rep{10}$
representations for $\Vt$ satsifying equations~\eqref{7} and~\eqref{9} is 
$h^1\left( \Xt,{\wedge}^{2}\Vt \right)=4$.
A class of such bundles was presented and shown to give rise to the exact MSSM 
spectrum at low-energy.  In particular, the spectrum had a \emph{single}, 
vector-like pair of Higgs superfields.  In this paper, we want 
to enlarge the low-energy theory to include a second pair of 
Higgs fields.  That is, we continue to constrain $\Vt$ to satsify 
eqns.~\eqref{7} and~\eqref{9}, but enlarge $h^1\left( \Xt,{\wedge}^{2}\Vt \right)$. 
As we discuss below, one class of bundles $\Vt$ leading to precisely two
vector-like pairs of Higgs superfields satsifies
\begin{equation}
  h^1\left( \Xt,{\wedge}^{2}\Vt \right)=10
  \,.
  \label{10}
\end{equation}
The bundles are similar to those presented in~\cite{hsm2}, differing essentially
in one of the two ideal sheaves involved in the construction.  

We now present a stable vector bundle $\Vt$ satisfying
constraints eqns.~\eqref{7},~\eqref{9} and~\eqref{10}. 
This is constructed as an extension 
\begin{equation}
  0 
  \longrightarrow
  \V1
  \longrightarrow
  \Vt
  \longrightarrow
  \V2
  \longrightarrow
  0
  \label{C}
\end{equation}
of two rank $2$ bundles, $\V1$ and $\V2$. Each of these is the tensor
product of a line bundle with a rank $2$ bundle pulled back from a
$\dP9$ factor of $\Xt$. Using the two projection maps, we
define
\begin{equation}
  \V1 =
  \oXt(-\tau_1+\tau_2) \otimes \p1^\ast(\W1)
  \,, \quad
  \V2 =
  \oXt(\tau_1-\tau_2) \otimes \p2^\ast(\W2)
  \,,
  \label{D}
\end{equation}
where
\begin{equation}
{\text span}\{\tau_{1},\tau_{2},\phi\}= H_{2}(\Xt,\mathbb{C})^{\ZZZ}
\label{burt1}
\end{equation}
is the $\ZZZ$ invariant part of the Kahler moduli space.
The two bundles, $\W1$ on $\B1$ and $\W2$ on $B_2$, are constructed
via an equivariant version of the Serre construction as
\begin{equation}
  0 
  \longrightarrow
  \chi_2^2
  \oB1(- f_1) 
  \longrightarrow
  \W1
  \longrightarrow
  \chi_2
  \oB1( f_1) \otimes I_3^{B_1}
  \longrightarrow
  0
  \
  \label{A}
\end{equation}
and
\begin{equation}
  0 
  \longrightarrow
  \chi_2^2
  \oB2(-f_2) 
  \longrightarrow
  \W2
  \longrightarrow
  \chi_2
  \oB2(f_2) \otimes I_6^{B_2}
  \longrightarrow
  0    
  \,,
  \label{B}
\end{equation}
where $I_3^{B_1}$ and $I_6^{B_2}$ denote ideal sheaves of $3$ and $6$
points in $B_1$ and $B_2$ respectively. Characters $\chi_1$ and
$\chi_2$ are third roots of unity which generate the first and second
factors of $\ZZZ$.\footnote{See~\cite{hsm1a,hsm2} for our notation for line
  bundles $\oXt(\cdots)$, etc.}


Note that $V_{1}$, $V_{2}$, $W_{1}$, and $W_{2}$ in eqns.~\eqref{D},~\eqref{A}
and~\eqref{B} respectively are constructed in the same manner 
as in~\cite{hsm2}.  Indeed, the line bundles $\oXt(\mp(\tau_1-\tau_2))$ in eqns.~\eqref{D}
and the ideal sheaf $I_6^{B_2}$ are taken to be identical 
to those in the exact MSSM case.  However, in order for $\wedge^{2}\tv$ to
satisfy condition eqn.~\eqref{10}, the ideal sheaf $I_3^{B_1}$ 
must now be chosen differently, as we now discuss.

Satisfying eqn.~\eqref{7} requires that one use ideal sheaves of 9
points in total.  As in the exact MSSM bundles~\cite{hsm2}, we continues to 
distribute these points into two different ideal sheaves, $I_3^{B_1}$
and $I_6^{B_2}$ on $B_1$ and $B_2$ respectively.  Furthermore, 
$I_6^{B_2}$ is chosen to be identical to that in~\cite{hsm2}, namely, the ideal
sheaf of the three fixed points of the second $\mathbb {Z}_{3}$ acting on $B_2$
taken with multiplicity 2.  However, to obtain $\wedge^{2}\tv$ satisfying 
eqn.~\eqref{10}, we now modify our choice of $I_3^{B_1}$. Note that there 
are four different choices of \ZZZ orbits of length 3 on $B_1$, and each
gives a different ideal sheaf of 3 points.  In~\cite{hsm2}, we took the three 
points to be the fixed points of the second $\mathbb {Z}_{3}$, which
are the singular points in the $3I_1$ Kodaira fibers.  To satisfy eqn.~\eqref{10},
however, we now define $I_3^{B_1}$ using the three fixed
points of the \emph{first} $\mathbb {Z}_{3}$ instead.  These all lie 
on the non-degenerate $T^2$ fiber over $0=[0:1]\subset\IP1$.
This allows one to obtain the MSSM spectrum with, additionally, 
a second pair of Higgs superfields.

We now extend the observable sector bundle $\V{}$ by adding a Wilson
line, $W$, with holonomy
\begin{equation}
  \mathrm{Hol}(W)=\ZZZ \subset Spin(10)
  \,.
  \label{12}
\end{equation}
The associated gauge connection spontaneously breaks $Spin(10)$ as
\begin{equation}
  Spin(10) \longrightarrow 
  SU(3)_{C} \times 
  SU(2)_{L} \times 
  U(1)_{Y} \times 
  U(1)_{B-L}
  \,,
  \label{13}
\end{equation}
where $SU(3)_{C} \times SU(2)_{L} \times U(1)_{Y}$ is the standard model gauge
group. Since $\ZZZ$ is Abelian and 
$\rank\big(Spin(10)\big)=5$, an additional rank one factor must appear.
For the chosen embedding of $\ZZZ$, 
this is precisely the gauged $B-L$ symmetry.

As discussed in~\cite{spec1}, the zero mode spectrum of $\V{} \oplus W$
on $X$ is determined as follows. Let $R$ be a representation of
$Spin(10)$, and denote the associated tensor product bundle of $\Vt$
 by $U_{R}(\Vt)$.  
Then, each sheaf
cohomology space $H^\ast\big(\Xt, U(\Vt)_{R}\big)$ carries a specific
representation of $\ZZZ$. Similarly, the Wilson line $W$ manifests
itself as a $\ZZZ$ group action on each representation $R$ of
$Spin(10)$.  As discussed in detail in~\cite{hsm1a,hsm2}, the low-energy
particle spectrum is given by

\begin{multline}
  \label{120} 
  \ker\big(\dslash_{V}\big) = \left( 
    H^0\big(\tx, \cO_{\tx}\big) \otimes \Rep{45} \right)^{\z3z3}
  \oplus \left( H^1\big(\tx, \tv^\ast \big) \otimes
    \barRep{16} \right)^{\z3z3} \\ 
  \oplus
  \left( H^1\big(\tx, \tv\big) \otimes
    \Rep{16} \right)^{\z3z3} \oplus 
  \left( H^1\big(\tx, \atv \big) \otimes \Rep{10} \right)^{\z3z3}
  \oplus 
  \left( H^1\big(\tx, \ad(\tv) \big) \otimes
    \Rep{1} \right)^{\z3z3}
  ,
\end{multline} 
where the superscript indicates the invariant subspace under the
action of $\ZZZ$.  The invariant cohomology space $\big( H^0(\tx,
\cO_{\tx}) \otimes \Rep{45} \big)^{\z3z3}$ corresponds to gauge
superfields in the low-energy spectrum carrying the adjoint
representation of the gauge group $SU(3)_{C} \times SU(2)_{L} 
\times U(1)_{Y} \times U(1)_{B-L}$. The matter cohomology spaces are
\begin{equation}
  \Big( H^1(\tx, \tv^\ast ) \otimes \barRep{16}\Big)^{\z3z3}
  ,~
  \Big( H^1(\tx, \tv) \otimes \Rep{16} \Big)^{\z3z3}
  ,~
  \Big( H^1(\tx, \atv ) \otimes \Rep{10} \Big)^{\z3z3}
  .
\end{equation}

First consider the $\barRep{16}$ representation. It follows from
eq.~\eqref{9} that no such representations occur. Hence, no $SU(3)_{C}
\times SU(2)_{L} \times U(1)_{Y} \times U(1)_{B-L}$ fields arising
from vector-like $\barRep{16}$-$\Rep{16}$ pairs appear in the
spectrum, as desired. Next, examine the $\Rep{16}$ representation. The
constraints~\eqref{7} and~\eqref{9} imply that
\begin{equation}
  h^1\left( \Xt,\Vt \right)=27
  \,.
  \label{14}
\end{equation}
One can calculate the $\ZZZ$ representation on
$H^1\big(\Xt,\Vt\big)$, as well as the Wilson line action on
$\Rep{16}$. We find that
\begin{equation}
H^1\big(\Xt,\Vt\big)=RG^{\oplus3},
\label{burt2}
\end{equation}
where $RG$ is the regular representation of $G=\ZZZ$ given by
\begin{equation}
RG=1 \oplus \chi_{1} \oplus \chi_{2} \oplus \chi_{1}^{2} \oplus \chi_{2}^{2}
\oplus \chi_{1}\chi_{2} \oplus \chi_{1}^{2}\chi_{2} \oplus 
\chi_{1}\chi_{2}^{2} \oplus \chi_{1}^{2}\chi_{2}^{2}.
\label{burt3}
\end{equation}
Furthermore, the Wilson line action can be chosen so that
\begin{multline}
  \label{burt4}
  \Rep{16}= \Big[
  \chi_{1}\chi_{2}^{2} \big(\Rep{3}, \Rep{2}, 1, 1 \big)
  \oplus \chi_{2}^{2} \big(\Rep{1},\Rep{1}, 6, 3 \big)
  \oplus \chi_{1}^{2}\chi_{2}^{2} \big(\barRep{3},\Rep{1}, -4, -1 \big)\Big]
  \oplus \\ \oplus 
  \Big[  \big(\Rep{1},\Rep{2}, -3, -3 \big) 
  \oplus \chi_{1}^{2} \big(\barRep{3},\Rep{1}, 2, -1 \big)\Big]
  \oplus \chi_{2} \big(\Rep{1},\Rep{1}, 0, 3 \big).
\end{multline}
Tensoring these together, we find that the invariant
subspace $\Big( H^1(\tx, \tv) \otimes \Rep{16} \Big)^{\z3z3}$ consists of three
families of quarks and leptons, each family
transforming as
\begin{equation}
  \ Q = \big(\Rep{3},   \Rep{2}, 1, 1 \big) \,,\quad
  \ u =\big(\barRep{3},\Rep{1}, -4, -1 \big) \,,\quad
  \ d =\big(\barRep{3},\Rep{1}, 2, -1 \big)
  \label{15}
\end{equation}
and
\begin{equation}
  \ L=\big(\Rep{1},\Rep{2}, -3, -3 \big) \,,\quad
  \ e=\big(\Rep{1},\Rep{1}, 6, 3 \big) \,,\quad
  \nu=\big(\Rep{1},\Rep{1}, 0, 3 \big)
  \label{16}
\end{equation}
under $SU(3)_{C} \times SU(2)_{L} \times U(1)_{Y} \times U(1)_{B-L}$.
We have displayed the quantum numbers $3Y$ and $3(B-L)$ for
convenience. Note from eqn.~\eqref{16} that each family contains a
right-handed neutrino, as desired.

Next, consider the $\Rep{10}$ representation. Recall from
eqn.~\eqref{10} that $h^1\big(\Xt,{\wedge}^{2}\Vt\big)=10$. We find
that the representation of $\ZZZ$
in $H^1\big(\Xt,{\wedge}^{2}\Vt\big)$ is given by  
\begin{equation}
  \ H^1 \big(\Xt,{\wedge}^{2}\Vt\big)= 
  \ ( \chi_1 + \chi_1^2 + \chi_{2} + \chi_{2}^{2})^{\oplus 2}
  \oplus \chi_{1}\chi_{2}^{2} \oplus \chi_{1}^{2}\chi_{2} 
  \,.
  \label{17}
\end{equation}
Furthermore, the Wilson line $W$ action is
\begin{equation}
  \Rep{10}= 
  \Big[
  \chi_1^2\big(\Rep{1},\Rep{2},3,0\big) \oplus 
  \chi_1^2\chi_2^2\big(\Rep{3},\Rep{1},-2,-2\big)
  \Big]
  \oplus 
  \Big[
  \chi_1\big(\Rep{1},\barRep{2},-3,0\big) \oplus 
  \chi_1\chi_2\big(\barRep{3},\Rep{1},2,2\big)
  \Big]
  \,.
  \label{19}
\end{equation}
Tensoring these actions together, one finds that the invariant
subspace of $(H^1\big(\Xt,{\wedge}^{2}\Vt\big) \otimes \Rep{10})^{\ZZZ}$ 
consists of two vector-like pairs, each pair transforming as 
\begin{equation}
  \ H_k=\big( \Rep{1},\Rep{2}, 3, 0 \big) \,,\quad
  \bar{H}_k=\big( \Rep{1},\barRep{2}, -3,  0 \big)
  \,,\quad  k=1,2 .
  \label{21}
\end{equation}
That is, there are two pairs of Higgs--Higgs conjugate fields
occurring as zero modes of our vacuum.

Finally, consider the $\Rep{1}$ representation of the $Spin(10)$ gauge
group. It follows from~\eqref{8}, the above discussion, and the fact
that the Wilson line action on $\Rep{1}$ is trivial that the number of
$\Rep{1}$ zero modes is given by the $\ZZZ$ invariant subspace of
$H^1\big(\Xt, \ad(\Vt) \big)$, which is denoted by
$H^1\big(\Xt, \ad(\Vt) \big)^\ZZZ$. Using the formalism
developed in~\cite{lss}, we find that
\begin{equation}
  h^1\left(\Xt, \ad(\Vt) \right)^\ZZZ
  =
  13
  \label{moduli}
\end{equation}
That is, there are $13$ vector bundle moduli.

Putting these results together, we conclude that the zero mode
spectrum of the observable sector has gauge group $SU(3)_C \times
SU(2)_L \times U(1)_Y \times U(1)_{B-L}$, contains three families of
quarks and leptons each with a right-handed neutrino, has \emph{two}
Higgs--Higgs conjugate pairs, and contains no exotic fields or
additional vector-like pairs of multiplets of any kind.  Furthermore,
there are $13$ vector bundle moduli.


\section{Cubic Yukawa Terms in the Superpotential}
\label{sec:couplings}

We now focus on computing Yukawa terms.  It follows
from eq.~\eqref{120} that the $4$-dimensional Higgs and quark/lepton
fields correspond to certain $\bar{\partial}$-closed $(0,1)$-forms on
$\Xt$ with values in the vector bundle $\wedge^2 \Vt$ and $\Vt$
respectively. Since both pairs of Higgs and Higgs-conjugate arise from the same
cohomology space, we will denote any of these $1$-forms simply as
$\Psi^{H}$. For the same reason, we schematically represent any
quark/lepton doublet by $\Psi^{(2)}$ and any singlet $1$-form by
$\Psi^{(1)}$, in any family. They can be written as
\begin{equation}
  \Psi^H = \psi^{H}_{\bar{\iota}[ab]}   
  , \diff \bar{z}^{\bar{\iota}}
  , \quad
  \Psi^{(1)}= \psi_{\bar{\iota}a}^{(1)}
  , \diff \bar{z}^{\bar{\iota}}
  , \quad
  \Psi^{(2)}= \psi_{\bar{\iota}b}^{(2)}
  , \diff \bar{z}^{\bar{\iota}}
  , 
\end{equation}
where $a$, $b$ are valued in the $SU(4)$ bundle $\Vt$ and
$\{z^\iota,\bar{z}^{\bar{\iota}}\}$ are coordinates on the Calabi-Yau
threefold $\Xt$. Doing the dimensional reduction of the
$10$-dimensional Lagrangian yields cubic terms in the superpotential
of the $4$-dimensional effective action. It turns
out~\cite{lssa} that the coefficients of the cubic couplings
are simply the various allowed ways to obtain a number out of the
forms $\Psi^H$, $\Psi^{(1)}$, $\Psi^{(2)}$. That is, schematically
\begin{equation}
  \label{eq:weq}
  W = \cdots + \lambda_{u}QHu + \lambda_{d}Q\bar{H}d
  +  \lambda_{\nu}LH\nu + \lambda_{e}L\bar{H}e 
\end{equation}
with the coefficients $\lambda$ determined by
\begin{equation}
  \label{eq:lambdaintegral}
  \begin{split}
    \lambda ~&= 
    \int_\Xt
    \Omega \wedge 
    \Tr\Big[ 
    \Psi^{(2)} \wedge \Psi^{H} \wedge \Psi^{(1)}
    \Big]
    = \\ &= 
    \int_\Xt
    \Omega \wedge 
    \Big( 
    \epsilon^{abcd}
    \psi^{(2)}_{\bar{\iota}a}
    \,
    \psi^{H}_{\bar{\kappa} [bc]} 
    \,
    \psi^{(1)}_{\bar{\epsilon}d}
    \Big)
    \diff \bar{z}^{\bar{\iota}} \wedge
    \diff \bar{z}^{\bar{\kappa}} \wedge
    \diff \bar{z}^{\bar{\epsilon}}
  \end{split}
\end{equation}
and $\Omega$ is the holomorphic $(3,0)$-form. Mathematically, we are
using the wedge product together with a contraction of the vector
bundle indices (that is, the determinant $\wedge^4\Vt = \oXt$) to
obtain a product
\begin{multline}
  \label{eq:cup}
  H^1\Big(\tx, \Vt \Big) \otimes 
  H^1\Big(\tx, \atv \Big) \otimes
  H^1\Big(\tx, \Vt \Big) 
  \longrightarrow \\ \longrightarrow
  H^3\Big(\tx, \Vt \otimes \atv \otimes \Vt\Big)
  \longrightarrow 
  H^3\Big(\tx, \cO_{\tx} \Big),
\end{multline}
plus the fact that on the Calabi-Yau manifold $\Xt$
\begin{equation}
  \label{eq:coho}
  H^3\Big(\Xt,\cO_{\Xt}\Big) =
  H^3\Big(\Xt, K_\Xt\Big) = 
  H_{\bar{\partial}}^{3,3}\Big(\Xt\Big) =
  H^6\Big(\Xt\Big)
\end{equation}
can be integrated over. If one were to use the heterotic string with
the ``standard embedding'', then the above product would simplify
further to the intersection of certain cycles in the Calabi-Yau
threefold. However, in our case
there is no such description.

Hence, to compute Yukawa terms, we must first analyze the cohomology
groups
\begin{equation} 
  H^1\Big(\tx, \Vt \Big),~
  H^1\Big(\tx, \atv \Big),~
  H^3\Big(\tx, \cO_{\tx} \Big)
  \label{140}
\end{equation}
and the action of $\z3z3$ on these spaces. We then have to evaluate
the product in eq.~\eqref{eq:cup}. As we will see in the following
sections, the two independent elliptic fibrations of $\Xt$ will force
some, but not all, products to vanish.

\subsection{The First Elliptic Fibration}
\label{sec:LSS1}

\subsubsection{The Leray Spectral Sequence}

As discussed in detail in~\cite{hsm1a,hsm2, lss,lssa}, the cohomology spaces on
$\tx$ are obtained by using two Leray spectral sequences.  In this
section, we consider the first of these sequences corresponding to the
projection
\begin{equation} 
  \tx \stackrel{\pi_{2}}{\longrightarrow} B_2.
  \label{150}
\end{equation}
For any sheaf $\cal{F}$ on $\tx$, the Leray spectral sequence tells us
that
\begin{equation} 
  H^i\Big( \tx, {\cal F} \Big) = \bigoplus^{p+q=i}_{p,q}
  H^p\Big(B_2, R^q\pi_{2\ast}\cF\Big),
  \label{160}
\end{equation}
where the only non-vanishing entries are for $p=0,1,2$ (since $\dim_\C
(B_2)=2$) and $q=0,1$ (since the fiber of $\tx$ is an elliptic curve,
therefore of complex dimension one). Note that the cohomologies
$H^{p}(B_2, R^q\pi_{2\ast}\cF)$ fill out the $2\times 3$
tableau\footnote{Recall that the zero-th derived push-down is just the
  ordinary push-down, $R^0\pi_{2\ast}=\pi_{2\ast}$.}
\begin{equation}
  \label{170} 
  \vcenter{ \def\w{34mm} \def\h{8mm} \xymatrix@C=0mm@R=0mm{
      {\scriptstyle q=1} & 
      *=<\w,\h>[F]{ H^0\big(B_2, R^1\pi_{2\ast}\cF\big) } &
      *=<\w,\h>[F]{ H^1\big(B_2, R^1\pi_{2\ast}\cF\big) } & 
      *=<\w,\h>[F]{ H^2\big(B_2, R^1\pi_{2\ast}\cF\big) } 
      \\ {\scriptstyle q=0} & 
      *=<\w,\h>[F]{ H^0\big(B_2, \pi_{2\ast}\cF\big) } & 
      *=<\w,\h>[F]{ H^1\big(B_2, \pi_{2\ast}\cF\big) } & 
      *=<\w,\h>[F]{ H^2\big(B_2, \pi_{2\ast}\cF\big) } 
      \\ & {\scriptstyle p=0} & {\scriptstyle p=1} & 
      {\scriptstyle p=2} 
    }} 
  \Rightarrow 
  H^{p+q}\Big(\Xt, \cF\Big)
  ,
\end{equation}
where ``$\Rightarrow H^{p+q}\big(\Xt, \cF\big)$'' reminds us of which
cohomology group the tableau is computing. Such tableaux are very
useful in keeping track of the elements of Leray spectral sequences.
As is clear from eq.~\eqref{160}, the sum over the diagonals yields the
desired cohomology of $\cF$. In the following, it will be very helpful
to define
\begin{equation} 
  H^p\Big(B_2, R^q\pi_{2\ast}\cF\Big) \equiv 
  \Hpq{\cF}{p}{q}
  .
  \label{180}
\end{equation}
Using this abbreviation, the tableau eq.~\eqref{170} reads
\begin{equation}
  \label{eq:Hpqtableau} 
  \vcenter{ \def\w{20mm} \def\h{8mm} \xymatrix@C=0mm@R=0mm{
      {\scriptstyle q=1} & 
      *=<\w,\h>[F]{ \Hpq{\cF}{0}{1} } &
      *=<\w,\h>[F]{ \Hpq{\cF}{1}{1} } &
      *=<\w,\h>[F]{ \Hpq{\cF}{2}{1} } 
      \\ {\scriptstyle q=0} & 
      *=<\w,\h>[F]{ \Hpq{\cF}{0}{0} } &
      *=<\w,\h>[F]{ \Hpq{\cF}{1}{0} } &
      *=<\w,\h>[F]{ \Hpq{\cF}{2}{0} } 
      \\ & {\scriptstyle p=0} & {\scriptstyle p=1} & 
      {\scriptstyle p=2} 
    }} 
  \Rightarrow 
  H^{p+q}\Big(\Xt, \cF\Big)
  .
\end{equation}

On the level of differential forms, we can understand the Leray
spectral sequence as decomposing differential forms into the number
$p$ of legs in the direction of the base and the number $q$ of legs in
the fiber direction. Obviously, this extra grading is preserved under
the wedge-product of the differential forms. Hence, any product
\begin{equation}
  H^i\Big( \Xt, \Fsheaf_1 \Big) \otimes
  H^j\Big( \Xt, \Fsheaf_2 \Big) 
  \longrightarrow
  H^{i+j}\Big( \Xt, \Fsheaf_1 \otimes \Fsheaf_2 \Big) 
\end{equation}
not only has to end up in overall degree $i+j$, but also has to
preserve the $(p,q)$-grading. That is,
\begin{equation}
  \label{eq:simpleproduct}
  \vcenter{\xymatrix@R=3ex{
      \Hpq{\Fsheaf_1}{p_1}{q_1}
      \ar@{}[r]|{\displaystyle \otimes} & 
      \Hpq{\Fsheaf_2}{p_2}{q_2}
      \ar[r] &
      \Hpq{\Fsheaf_1\otimes\Fsheaf_2}{p_1+p_2}{q_1+q_2}
      \\
      H^{p_1+q_1}\Big( \Xt, \Fsheaf_1 \Big) 
      \ar@{}[u]|{\displaystyle \cap}
      \ar@{}[r]|{\displaystyle \otimes} & 
      H^{p_2+q_2}\Big( \Xt, \Fsheaf_2 \Big) 
      \ar@{}[u]|{\displaystyle \cap}
      \ar[r] &
      H^{p_1+p_2+q_1+q_2}\Big( \Xt, \Fsheaf_1 \otimes \Fsheaf_2 \Big) 
      \ar@{}[u]|{\displaystyle \cap}
      \,.
    }}
\end{equation}
This will be used in the following discussion.

\subsubsection{The First Leray Decomposition of the Volume Form}

Let us first discuss the $(p,q)$ Leray tableau for the sheaf
$\cF=\cO_{\tx}$, which is the last term in eq.~\eqref{140}. Since this
is the trivial line bundle, it immediately follows that
\begin{equation}
  \label{190} 
  \vcenter{ \def\w{20mm} \def\h{6mm} \xymatrix@C=0mm@R=0mm{
      {\scriptstyle q=1} & 
      *=<\w,\h>[F]{ 0 } & *=<\w,\h>[F]{ 0 } &
      *=<\w,\h>[F]{ \Rep{1} } 
      \\ {\scriptstyle q=0} & 
      *=<\w,\h>[F]{ \Rep{1} } & 
      *=<\w,\h>[F]{ 0 } & 
      *=<\w,\h>[F]{ 0 } 
      \\ & 
      {\scriptstyle p=0} & {\scriptstyle p=1} & {\scriptstyle p=2} 
    }} 
  \Rightarrow
  H^{p+q}\Big(\tx,\cO_\tx\Big)
  .
\end{equation}
From eqns.~\eqref{160} and~\eqref{190} we see that
\begin{equation} 
  H^3 \Big( \tx, \cO_\tx \Big) = \Hpq{\cO_\tx}{2}{1} = \Rep{1}
  ,
  \label{200}
\end{equation}
where the $\Rep{1}$ indicates that $H^3 ( \tx, \cO_{\tx} )$ is a
one-dimensional space carrying the trivial action of $\z3z3$.

\subsubsection{The First Leray Decomposition of Higgs Fields}

Now consider the $(p,q)$ Leray tableau for the sheaf $\cF = \atv$,
which is the second term in eq.~\eqref{140}. This can be explicitly
computed and is given by 
\begin{equation}
  \label{22} 
  \vcenter{ 
    \def\w{20mm} 
    \def\ww{50mm} 
    \def\h{16mm} 
    \xymatrix@C=0mm@R=0mm{
      {\scriptstyle q=1} & 
      *=<\w,\h>[F]{ \chi_2\oplus\chi_2^2 } & 
      *=<\ww,\h>[F]{ 
        \begin{array}{c}
          2 (\chi_1\oplus\chi_1^2) \oplus 2 (\chi_2\oplus\chi_2^2)
          \\
          \oplus \chi_1 \chi_2^2 \oplus \chi_1^2 \chi_2
        \end{array}
      } & 
      *=<\w,\h>[F]{ 0 } 
      \\ {\scriptstyle q=0} & 
      *=<\w,\h>[F]{ 0 } &
      *=<\ww,\h>[F]{
        \begin{array}{c}
          2 (\chi_1\oplus\chi_1^2) \oplus \chi_2\oplus\chi_2^2
          \\
          \oplus \chi_1 \chi_2^2 \oplus \chi_1^2 \chi_2
        \end{array}
      } & 
      *=<\w,\h>[F]{ 0 } \\ &
      {\scriptstyle p=0} & {\scriptstyle p=1} & {\scriptstyle p=2} 
    }} 
  \Rightarrow 
  H^{p+q}\Big(\Xt, \atv\Big)
  .
\end{equation}
In general, it follows from eq.~\eqref{eq:Hpqtableau} that $H^1(\tx,
\atv )$ is the sum of the entries on the first diagonal,
\begin{equation} 
  \begin{split}
    H^1\Big(\tx, \atv \Big) 
    =&\; 
    \Hpq{\atv}{0}{1} \oplus \Hpq{\atv}{1}{0} 
    \\
    =&\;
    2 \big( \chi_1\oplus\chi_1^2 \oplus \chi_2\oplus\chi_2^2 \big)
    \oplus \chi_1 \chi_2^2 \oplus \chi_1^2 \chi_2
    .
  \end{split}
  \label{eq:H1atv}
\end{equation}

\subsubsection{The First Leray Decomposition of the Quark/Lepton Fields}

Now consider the $(p,q)$ Leray tableau for the sheaf $\cF = \Vt$,
which is the first term in eq.~\eqref{140}. This can be explicitly
computed and is given by
\begin{equation}
  \label{26a} 
  \vcenter{ \def\w{20mm} \def\h{6mm} \xymatrix@C=0mm@R=0mm{
      {\scriptstyle q=1} & 
      *=<\w,\h>[F]{ RG^{\oplus2} } & 
      *=<\w,\h>[F]{ 0 } & 
      *=<\w,\h>[F]{ 0 } 
      \\ {\scriptstyle q=0} & 
      *=<\w,\h>[F]{ 0 } &
      *=<\w,\h>[F]{ RG } & 
      *=<\w,\h>[F]{ 0 } \\ &
      {\scriptstyle p=0} & {\scriptstyle p=1} & {\scriptstyle p=2} 
    }} 
  \Rightarrow 
  H^{p+q}\Big(\Xt, \Vt\Big)
  ,
\end{equation}
where $RG$ is the regular representation of $\z3z3$ given by
\begin{equation} 
  RG=
  1 \oplus \chi_{1} \oplus \chi_{2} \oplus \chi_{1}^{2} \oplus 
  \chi_{2}^{2} \oplus \chi_{1}\chi_{2} \oplus \chi_{1}
  \chi_{2}^{2} \oplus \chi_{1}^{2} \chi_{2} \oplus \chi_{1}^{2} \chi_{2}^{2}. 
  \label{27a}
\end{equation}
It follows from eq.~\eqref{eq:Hpqtableau} that $H^1(\tx, \Vt )$ is the sum of the
two subspaces
\begin{equation} 
  H^1\Big(\tx, \Vt \Big) = \Hpq{\Vt}{0}{1} \oplus  \Hpq{\Vt}{1}{0}
  .
  \label{28a}
\end{equation}
Furthermore, eq.~\eqref{26a} tells us that
\begin{equation}
  \Hpq{\Vt}{0}{1}= RG^{\oplus2}
  , \quad 
  \Hpq{\Vt}{1}{0} = RG.
  \label{29a}
\end{equation}
Technically, the structure of eq.~\eqref{28a} is associated with the
fact that the cohomology $H^{\ast}\big(\Xt,\Vt\big)$ decomposes into
$H^{\ast}\big(\Xt,V_{1}\big) \oplus H^{\ast}\big(\Xt,V_{2}\big)$. It
turns out that the two subspaces in eq.~\eqref{28a} arise as
\begin{equation}
  RG=H^{1}\big(\Xt,V_{1}\big), \quad RG^{\oplus2}=H^{1}\big(\Xt,V_{2}\big)
  \label{hope4}
\end{equation}
respectively.

\subsubsection{The (p,q) Selection Rule}

Having computed the decompositions of $H^3(\tx, \cO_{\tx} )$,
$H^1(\tx, \atv)$ and $H^1(\tx, \Vt )$ into their ${(p,q)}$ Leray
subspaces, we can now analyze the $(p,q)$ components of the triple
product
\begin{equation}
  \label{eq:product}
  H^1\Big(\tx, \Vt \Big) \otimes 
  H^1\Big(\tx, \atv \Big) \otimes
  H^1\Big(\tx, \Vt \Big) 
  \longrightarrow 
  H^3\Big(\tx, \cO_{\tx} \Big)  
\end{equation}
given in eq.~\eqref{eq:cup}. Inserting eqns.~\eqref{eq:H1atv}
and~\eqref{28a}, we see that
\begin{multline}
  \label{eq:pqCup}
  H^1\Big(\tx, \Vt \Big) \otimes 
  H^1\Big(\tx, \atv \Big) \otimes
  H^1\Big(\tx, \Vt \Big) 
  = \\ 
  \Big( \Hpq{\Vt}{0}{1} \oplus 
  \Hpq{\Vt}{1}{0} \Big)
  \otimes
  \Hpq{\atv}{1}{0} 
  \otimes 
  \Big( \Hpq{\Vt}{0}{1} \oplus 
  \Hpq{\Vt}{1}{0} \Big)
  = \\ 
  \underbrace{\Big( 
    {\scriptstyle 
      \Hpq{\Vt}{0}{1} \otimes
      \Hpq{\atv}{1}{0} \otimes 
      \Hpq{\Vt}{1}{0} 
    }
    \Big)^{\oplus2} 
  }_{\text{total $(p,q)$ degree }=\, (2,1)}
  \oplus
  \underbrace{\Big( 
    {\scriptstyle 
      \Hpq{\Vt}{1}{0} \otimes
      \Hpq{\atv}{1}{0} \otimes 
      \Hpq{\Vt}{1}{0} 
    }
    \Big) 
  }_{\text{total $(p,q)$ degree }=\, (3,0)}
  \oplus
  \underbrace{\Big( 
    {\scriptstyle 
      \Hpq{\Vt}{0}{1} \otimes
      \Hpq{\atv}{0}{1} \otimes 
      \Hpq{\Vt}{0}{1} 
    }
    \Big) 
  }_{\text{total $(p,q)$ degree }=\, (0,3)}
\end{multline}
Because of the $(p,q)$ degree, we see from eq.~\eqref{200} that only
the first term can have a non-zero product in
\begin{equation}
  \label{48}
  H^3 \Big( \tx, \cO_\tx \Big) = \Hpq{\cO_\tx}{2}{1}
  .
\end{equation}
It follows that the first quark/lepton family, which arises from
\begin{equation} 
  \Hpq{\Vt}{1}{0} = RG,
  \label{49}
\end{equation}
will form non-vanishing Yukawa terms with the second and third
quark/lepton families coming from
\begin{equation} 
  \Hpq{\Vt}{0}{1} = RG^{\oplus2}.
  \label{49a}
\end{equation}
All other Yukawa couplings must vanish. We refer to this as the
\emph{$(p,q)$ Leray degree selection rule}. We conclude that the only
non-zero product in eq.~\eqref{eq:product} is of the form
\begin{equation}
  \label{eq:pqproduct}
  \Hpq{\Vt}{0}{1} \otimes
  \Hpq{\atv}{1}{0} \otimes 
  \Hpq{\Vt}{1}{0} 
  \longrightarrow
  \Hpq{\cO_\Xt}{2}{1}
  .
\end{equation}
Roughly what happens is the following. The holomorphic $(3,0)$-form
$\Omega$ has two legs in the base and one leg in the fiber direction.
According to eq.~\eqref{eq:H1atv}, both $1$-forms $\Psi^H$
corresponding to Higgs and Higgs conjugate have their one leg in the
base direction.  Therefore, the wedge product in
eq.~\eqref{eq:lambdaintegral} can only be non-zero if one quark/lepton
$1$-form $\Psi$ has its leg in the base direction and the other
quark/lepton $1$-form $\Psi$ has its leg in the fiber direction.

We conclude that due to a selection rule for the $(p,q)$ Leray degree,
the Yukawa terms in the effective low-energy theory can involve only a
coupling of the first quark/lepton family to the second and third.
All other Yukawa couplings must vanish.

\subsection{The Second Elliptic Fibration}
\label{sec:LSS2}

\subsubsection{The Second Leray Spectral Sequence}

So far, we only made use of the fact that our Calabi-Yau manifold is
an elliptic fibration over the base $B_2$. But the $\dP9$ surface
$B_2$ is itself elliptically fibered over $\IP1$. Consequently,
there is yet another selection rule coming from the second elliptic
fibration. Therefore, we now consider the second Leray spectral
sequence corresponding to the projection
\begin{equation} 
  B_2 \stackrel{\beta_{2}}{\longrightarrow}
  \IP1.
  \label{50}
\end{equation}
For any sheaf $\cFt$ on $B_2$, the Leray sequence now starts with a $2
\times 2$ Leray tableau
\begin{equation}
  \label{52} 
  \vcenter{ \def\w{40mm} \def\h{8mm} \xymatrix@C=0mm@R=0mm{
      {\scriptstyle t=1} & 
      *=<\w,\h>[F]{ H^0\big( \IP1, R^1\beta_{2\ast} \cFt\big) } & 
      *=<\w,\h>[F]{ H^1\big( \IP1, R^1\beta_{2\ast}\cFt\big) } \\ 
      {\scriptstyle t=0} &
      *=<\w,\h>[F]{ H^0\big( \IP1, \beta_{2\ast} \cFt \big) } &
      *=<\w,\h>[F]{ H^1\big( \IP1, \beta_{2\ast} \cFt \big) } \\ &
      {\scriptstyle s=0} & {\scriptstyle s=1} }} 
  \Rightarrow
  H^{s+t}\Big(B_2, \cFt \Big)
  .
\end{equation}
Again, the sum over the diagonals yields the desired cohomology of
$\cFt$. Note that to evaluate the product eq.~\eqref{eq:pqproduct}, we
need the $[s,t]$ Leray tableaux for
\begin{equation}
  \cFt = 
  R^1\pi_{2\ast} \big(\Vt \big)
  ,~
  \pi_{2\ast} \big(\Vt \big)
  ,~
  \pi_{2\ast} \big( \atv \big)
  ,~
  R^1\pi_{2\ast} \big(\cO_\Xt\big)
  .
\end{equation}
In the following, it will be useful to define
\begin{equation} 
  H^{s} \bigg( \IP1, 
  R^t \beta_{2*} \Big( R^q \pi_{2\ast}\big(\cF\big)  \Big) \bigg) 
  \equiv 
  \Hst{\cF}{q}{s}{t}
  .
  \label{53}
\end{equation}
One can think of $\Hst{\cF}{q}{s}{t}$ as the subspace of
$H^\ast\big(\Xt, \cF\big)$ that can be written as forms with $q$ legs
in the $\pi_2$-fiber direction, $t$ legs in the $\beta_2$-fiber
direction, and $s$ legs in the base $\IP1$ direction.

\subsubsection{The Second Leray Decomposition of the Volume Form}

Let us first discuss the $[s,t]$ Leray tableau for
$\cFt=R^1\pi_{2\ast} \big(\cO_{\tx}\big)=K_{B_2}$, the canonical line
bundle. It follows immediately that
\begin{equation}
  \label{54} 
  \vcenter{ \def\w{20mm} \def\h{6mm} \xymatrix@C=0mm@R=0mm{
      {\scriptstyle t=1} & *=<\w,\h>[F]{ 0 } & *=<\w,\h>[F]{ \Rep{1} }
      \\ {\scriptstyle t=0} & *=<\w,\h>[F]{ 0 } & *=<\w,\h>[F]{ 0 } \\ &
      {\scriptstyle s=0} & {\scriptstyle s=1} }} 
  \Rightarrow
  H^{s+t}\Big(B_2, R^1\pi_{2\ast} \big(\cO_{\tx}\big) \Big)  
  .
\end{equation}
In our notation, this means that
\begin{equation}
  H^2\Big(B_2, R^1\pi_{2\ast} \big(\cO_{\tx}\big)\Big) = 
  \Hst{\cO_\Xt}{1}{1}{1}
\end{equation}
has pure $[s,t]=[1,1]$ degree. To summarize, we see that
\begin{equation} 
  \label{55}
  H^3\Big( \Xt, \cO_\Xt \Big) = 
  \Hpq{\cO_\Xt}{2}{1} =
  \Hst{\cO_\Xt}{1}{1}{1} =
  \Rep{1}
  .
\end{equation}

\subsubsection{The Second Leray Decomposition of Higgs Fields}

Now consider the $[s,t]$ Leray tableau for the sheaf
$\cFt=\pi_{2\ast}\big(\atv\big)$.  This can be explicitly computed 
and is given by
\begin{equation}
  \label{57} 
  \vcenter{ \def\w{50mm} \def\h{8mm} \xymatrix@C=0mm@R=0mm{
      {\scriptstyle t=1} & 
      *=<\w,\h>[F]{ 
        \chi_{1} \oplus \chi_{1}^{2} \oplus 
        2 \chi_{2} \oplus \chi_{2}^{2}
        \oplus \chi_{1} \chi_{2}^{2} 
      } & 
      *=<\w,\h>[F]{ 0 } \\
      {\scriptstyle t=0} & 
      *=<\w,\h>[F]{ \chi_2\oplus\chi_2^2 } & 
      *=<\w,\h>[F]{ 
        \chi_{1} \oplus \chi_{1}^{2} \oplus 
        \chi_{2}^{2} \oplus \chi_{1}^{2} \chi_{2}}  
      \\ & {\scriptstyle s=0} &
      {\scriptstyle s=1} }} 
  \Rightarrow 
  H^{s+t}\Big(B_2, R^1\pi_{2\ast} \big(\atv\big) \Big)    
  .
\end{equation}
\begin{equation}
  \label{57a} 
  \vcenter{ \def\w{50mm} \def\h{8mm} \xymatrix@C=0mm@R=0mm{
      {\scriptstyle t=1} & 
      *=<\w,\h>[F]{ 
        \chi_{1} \oplus \chi_{1}^{2} 
        \oplus \chi_{2}
        \oplus \chi_{1} \chi_{2}^{2} 
      } & 
      *=<\w,\h>[F]{ 0 } \\
      {\scriptstyle t=0} & 
      *=<\w,\h>[F]{ 0 } & 
      *=<\w,\h>[F]{ 
        \chi_{1} \oplus \chi_{1}^{2} 
        \oplus \chi_{2}^{2} 
        \oplus \chi_{1}^{2} \chi_{2}
      }  
      \\ & {\scriptstyle s=0} &
      {\scriptstyle s=1} }} 
  \Rightarrow 
  H^{s+t}\Big(B_2, \pi_{2\ast} \big(\atv\big) \Big)    
  .
\end{equation}
This means that the $10$ copies of the $\Rep{10}$ of $Spin(10)$ given
in eq.~\eqref{eq:H1atv} split as
\begin{equation}
  \begin{split}
    H^1\Big( \Xt, \atv \Big) 
    =&\; 
    \Hpq{\atv}{1}{0} \oplus  \Hpq{\atv}{0}{1} 
    \\
    =&\;
    \Big( \Hst{\atv}{0}{0}{1} \oplus \Hst{\atv}{0}{1}{0} \Big)
    \oplus \Hst{\atv}{1}{0}{0}
    \label{beta}
  \end{split}
\end{equation}
where
\begin{equation}
  \label{58}
  \begin{split}
    \Hst{\atv}{0}{0}{1} ~&=
     \chi_1 \oplus \chi_1^2 \oplus 
     \chi_2 \oplus \oplus \chi_1 \chi_2^2, 
    \\
    \Hst{\atv}{0}{1}{0} ~&=    
     \chi_1 \oplus \chi_1^2 \oplus 
     \chi_2^2 \oplus \chi_1^2 \chi_2
     \\
    \Hst{\atv}{1}{0}{0} ~&=
     \chi_2 \oplus \chi_2^2
     .
  \end{split}
\end{equation}
Note that
\begin{equation} 
  \begin{split}
    H^1\Big(\tx, \atv \Big) 
    =&\;
    \Hst{\atv}{0}{0}{1} \oplus \Hst{\atv}{0}{1}{0} 
    \oplus \Hst{\atv}{1}{0}{0}
    \\
    =&\;
    2 \big( \chi_1\oplus\chi_1^2 \oplus \chi_2\oplus\chi_2^2 \big)
    \oplus \chi_1 \chi_2^2 \oplus \chi_1^2 \chi_2
    ,
  \end{split}
  \label{60}
\end{equation}
see eq.~\eqref{eq:H1atv}.

\subsubsection{The Second Leray Decomposition of the Quark/Lepton Fields}

Finally, let us consider the $[s,t]$ Leray tableau for the
quark/lepton fields. We have already seen that, due to the $(p,q)$
selection rule, both the first quark/lepton family arising from
\begin{equation}  
  \Hpq{\Vt}{1}{0} = RG
  \label{AA}
\end{equation}
and the second and third quark/lepton families coming from
\begin{equation}   
  \Hpq{\Vt}{0}{1} = RG^{\oplus2} 
  \label{BB}
\end{equation}
must occur in non-vanishing Yukawa interactions.  Therefore, we are
only interested in the $[s,t]$ decomposition of each of these
subspaces. The $ \Hpq{\Vt}{0}{1}$ subspace is associated with the
degree $0$ cohomology of the sheaf $R^1\pi_{2\ast} \big( \Vt \big)$.
The corresponding Leray tableau is given by
\begin{equation}
  \label{61} 
  \vcenter{ \def\w{20mm} \def\h{6mm} \xymatrix@C=0mm@R=0mm{
      {\scriptstyle t=1} & 
      *=<\w,\h>[F]{ 0 } & 
      *=<\w,\h>[F]{ 0 } \\
      {\scriptstyle t=0} & 
      *=<\w,\h>[F]{ RG^{\oplus2} } & 
      *=<\w,\h>[F]{ 0 }
      \\ & {\scriptstyle s=0} & {\scriptstyle s=1} }} 
  \Rightarrow
  H^{s+t}\Big(B_2, R^1\pi_{2\ast} \big(\Vt \big)
  \Big)     
  .
\end{equation}
It follows that the second and third families of quarks/leptons has
$[s,t]$ degree $[0,0]$,
\begin{equation} 
  \label{62}
  \Hpq{\Vt}{0}{1} = 
  \Hst{\Vt}{1}{0}{0}= 
  RG^{\oplus2}
  .
\end{equation}
The $ \Hpq{\Vt}{1}{0}$ subspace is associated with the degree $1$
cohomology of the sheaf $\pi_{2\ast} \big( \Vt \big)$. The
corresponding Leray tableau is given by
\begin{equation}
  \label{61a} 
  \vcenter{ \def\w{20mm} \def\h{6mm} \xymatrix@C=0mm@R=0mm{
      {\scriptstyle t=1} & 
      *=<\w,\h>[F]{ RG } & 
      *=<\w,\h>[F]{ 0 } \\
      {\scriptstyle t=0} & 
      *=<\w,\h>[F]{ 0 } & 
      *=<\w,\h>[F]{ 0 }
      \\ & {\scriptstyle s=0} & {\scriptstyle s=1} }} 
  \Rightarrow
  H^{s+t}\Big(B_2,\pi_{2\ast} \big(\Vt \big)
  \Big)     
  .
\end{equation}
It follows that the first family of quarks/leptons has
$[s,t]$ degree $[0,1]$,
\begin{equation} 
  \label{62a}
  \Hpq{\Vt}{1}{0} = 
  \Hst{\Vt}{0}{0}{1}= 
  RG
  .
\end{equation}

\subsubsection{The [s,t] Selection Rule}

Having computed the decompositions of the relevant cohomology spaces
into their $[s,t]$ Leray subspaces, we can now calculate the triple
product eq.~\eqref{eq:cup}. The $(p,q)$ selection rule dictates that
the only non-zero product is of the form eq.~\eqref{eq:pqproduct}. Now
split each term in this product into its $[s,t]$ subspaces, as given
in eqns.~\eqref{55},~\eqref{beta},~\eqref{58},~\eqref{62} and~\eqref{62a}. The
result is
\begin{multline} 
  \label{65}
  \Hst{\Vt}{1}{0}{0} 
  \otimes
  \Big( \Hst{\atv}{0}{0}{1} \oplus \Hst{\atv}{0}{1}{0} \Big)
  \otimes 
  \Hst{\Vt}{0}{0}{1}
  \longrightarrow
  \Hst{\cO_\Xt}{1}{1}{1} 
  .
\end{multline}
Clearly, this triple product vanishes by degree unless we choose the
$\Hst{\atv}{0}{1}{0}$ from the $\Hpq{\atv}{1}{0}$ subspace.  In this
case, eq.~\eqref{65} becomes
\begin{equation} 
  \label{65-1}
  \Hst{\Vt}{1}{0}{0} 
  \otimes
  \Hst{\atv}{0}{1}{0}
  \otimes
  \Hst{\Vt}{1}{0}{1}
  \longrightarrow
  \Hst{\cO_\Xt}{1}{1}{1} 
  ,
\end{equation}
which is consistent.

We conclude that there is, in addition to the $(p,q)$ selection rule
discussed above, an \emph{$[s,t]$ Leray degree selection rule}. This
rule continues to allow non-vanishing Yukawa couplings of the first
quark/lepton family with the second and third quark/lepton families,
but only through the
\begin{equation}
  \Hst{\atv}{0}{1}{0}= \chi_{1} \oplus \chi_{1}^{2} \oplus 
  \chi_{2}^{2} \oplus \chi_{1}^{2}\chi_{2}
  \label{alpha}
\end{equation}
component of $\Hpq{\atv}{1}{0}$ in eq.~\eqref{beta}.

\subsubsection{Wilson Lines}

We have, in addition to the $SU(4)$ instanton, a non-vanishing Wilson
line. Its effect is to break the $Spin(10)$ gauge group down to the
desired $SU(3)_{C} \times SU(2)_{L} \times U(1)_{Y} \times U(1)_{B-L}$
gauge group.  First, consider the $\Rep{16}$ matter representations.
We choose the Wilson line $W$ so that its $\z3z3$ action on each
$\Rep{16}$ is given by
\begin{equation}
  \Rep{16}= \Big[
  \chi_{1}\chi_{2}^{2} Q
  \oplus \chi_{2}^{2} e
  \oplus \chi_{1}^{2}\chi_{2}^{2} u \Big]
  \oplus 
  \Big[ L
  \oplus \chi_{1}^{2} d \Big]
  \oplus \chi_{2} \nu,
  \label{hope1}
\end{equation}
where the representations $Q$,$u$,$d$ and $L$,$\nu$,$e$ were defined
in eqns.~\eqref{15} and~\eqref{16}, respectively. Recall from
eqns.~\eqref{28a} and~\eqref{29a} that $H^{1}\big( \Xt, \Vt \big)=RG
\oplus RG^{\oplus2}$.  Tensoring any $RG$ subspace of the cohomology
space $H^{1}\big(\Xt,\Vt\big)$ with a $\Rep{16}$ using
eqns.~\eqref{27a} and~\eqref{hope1}, we find that the invariant
subspace under the $\z3z3$ action is
\begin{equation}
  \Big( RG \otimes \Rep{16} \Big)^{\z3z3}= \Span \big\{Q,u,d,L,\nu,e \big\}.
  \label{hope2}
\end{equation}
It follows that each $RG$ subspace of $H^{1}\big(\Xt,\Vt\big)$
projects to a complete quark/lepton family at low-energy. This
justifies our identification of the subspace $RG$ with the first
quark/lepton family and the subspace $RG^{\oplus2}$ with the second
and third quark/lepton families throughout the text.

Second, notice that each fundamental matter field in the $\Rep{10}$
can be broken to a Higgs field, a color triplet, or projected out. In
particular, we are going to choose the Wilson line $W$ so that its
$\z3z3$ action on a $\Rep{10}$ representation of $Spin(10)$ is given
by
\begin{equation} 
  \Rep{10}=
  \Big[ 
  \chi_{1}^{2} H \oplus \chi_{1}^{2} \chi_{2}^{2} C
  \Big] \oplus \Big[ 
  \chi_{1} \bar{H} \oplus \chi_{1} \chi_{2}
  \bar{C}
  \Big].
  \label{66}
\end{equation}
From eqn.~\eqref{19}, we see that $H$ and $\bar{H}$ are the Higgs and
Higgs conjugate representations 
\begin{equation}
   \label{higgsrep}
   H=(\Rep{1},\Rep{2},3,0), \quad \bar{H}=(\Rep{1},\barRep{2},-3,0)
\end{equation}
and
\begin{equation}  
  C=\big( \Rep{3},\Rep{1}, -2, -2 \big),\quad 
  \bar{C}=\big( \barRep{3}, \Rep{1}, 2, 2 \big)
  \label{67}
\end{equation}
are the color triplet representations of $SU(3)_{C} \times SU(2)_{L}
\times U(1)_{Y} \times U(1)_{B-L}$.  Tensoring this with the
cohomology space $H^1\big(\Xt,\atv\big)$, we find the invariant
subspace under the combined $\z3z3$ action on the cohomology 
space, eqns.~\eqref{beta},~\eqref{58}, and
the Wilson line eqn.~\eqref{66}, to be
\begin{equation}
  \Big( H^1\big(\Xt,\atv\big) \otimes \Rep{10} \Big)^\z3z3
  = 
  \Span
  \big\{H_{1},\bar{H}_{1},H_{2},\bar{H}_{2}\big\}
  . 
\end{equation}
Note that $H_{1},\bar{H}_{1},H_{2},\bar{H}_{2}$ each arise from a
different $\Rep{10}$ representation.  The pairing $H_{k}$,
$\bar{H}_{k}$ for $k=1,2$ will be explained below.  Therefore, as
stated in eq.~\eqref{21}, precisely two pairs of Higgs--Higgs
conjugate fields survive the $\z3z3$ quotient. As required for any
realistic model, all color triplets are projected out.  The new
information now is the $(p,q)$ and $[s,t]$ degrees of the Higgs
fields. Using the decompositions eqns.~\eqref{eq:H1atv}, \eqref{beta},
and~\eqref{58} of $H^1\big(\Xt,\atv\big)$, we find
\begin{multline}
  \Big( H^1\big(\Xt,\atv\big) \otimes \Rep{10} \Big)^{\z3z3}
  =
  \Big( \Hpq{\atv}{1}{0} \otimes \Rep{10} \Big)^{\z3z3}
  = \\ =
  \underbrace{
    \Big( \Hst{\atv}{0}{0}{1} \otimes \Rep{10} \Big)^\z3z3
  }_{= span \lbrace H_{2}, \bar{H}_{2} \rbrace\ }
  \oplus 
  \underbrace{
    \Big( \Hst{\atv}{0}{1}{0} \otimes \Rep{10} \Big)^\z3z3
  }_{= span \lbrace H_{1}, \bar{H}_{1} \rbrace }
  .
\end{multline}
The dimensions and basis of the two terms on the right side of this
expression are determined by taking the tensor product of
eqns.~\eqref{58} and~\eqref{66} and keeping the $\z3z3$ invariant
part. Note that the subspace forming the non-zero Yukawa couplings in
eq.~\eqref{65-1}, namely $\Hst{\atv}{0}{1}{0}$, projects to only one 
of the two Higgs--Higgs conjugate pairs in the low-energy theory.

We label this pair as $H_{1}$, $\bar{H}_{1}$, despite the fact that
they arise from different $\Rep{10}$ representations.  The remaining
pair we denote as $H_{2}$, $\bar{H}_{2}$.  Since these are projected
from the $\Hst{\atv}{0}{0}{1}$ subspace, they are forbidden from
forming \emph{cubic} Yukawa couplings with quarks/leptons. To
conclude, of the two Higgs-Higgs conjugate pairs $(H_{k},\bar{H}_{k}),
k=1,2$ in the low-energy spectrum, only $(H_{1},\bar{H}_{1})$ can form
non-zero cubic Yukawa couplings.  Such couplings are disallowed for
$(H_{2},\bar{H}_{2})$ by the ``stringy'' $[s,t]$ Leray selection rule.

\subsection{Yukawa Couplings}
\label{sec:Yukawa}

We have analyzed cubic terms in the superpotential of the form
\begin{equation} 
  \lambda_{u,ij}^{k} Q_{i} H_{k} u_{j},  \quad 
  \lambda_{d,ij}^{k} Q_{i} \bar{H}_{k} d_{j}, \quad 
  \lambda_{\nu,ij}^{k} L_{i} H_{k} \nu_{j},  \quad 
  \lambda_{e,ij}^{k} L_{i} \bar{H}_{k} e_{j} 
  \label{67-1}
\end{equation}
where
\begin{itemize}
\item each coefficient $\lambda$ is determined by an integral
  of the form of eq.~\eqref{eq:lambdaintegral},
\item $Q_i$,$L_{i}$ for $i=1,2,3$ are the electroweak doublets of the three
  quarks/lepton families respectively,
\item $u_{j}$,$d_{j}$,$\nu_{j}$,$e_{j}$ for $j=1,2,3$ are the
  electroweak singlets of the three quark/lepton families
  respectively,
\item $H_{k}, k=1,2$ are the Higgs fields, and
\item $\bar{H}_{k}, k=1,2$ are the Higgs conjugate fields.
\end{itemize}
We found that they are subject to two independent selection rules
coming from the two independent torus fibrations. The first selection
rule is that the total $(p,q)$ degree is $(2,1)$.  Since the $(p,q)$
degrees for the first quark/lepton family, the second and third
quark/lepton families and all the Higgs fields are $(0,1)$, $(1,0)$ and
$(1,0)$ respectively, it follows that the only non-vanishing $\lambda$
coefficients allowed by the $(p,q)$ selection rules are of the form
\begin{equation}
  \lambda_{u,1j}^{k}, \lambda_{u,j1}^{k} \quad \lambda_{d,1j}^{k}, \lambda_{d,j1}^{k}\quad 
  \lambda_{\nu,1j}^{k}, \lambda_{\nu,j1}^{k}
  \quad \lambda_{e,1j}^{k}, \lambda_{e,j1}^{k}
  \label{gamma2}
\end{equation}
for $j=2,3$ and $k=1,2$. That is, the only non-zero Yukawa terms couple the first
family to the second and third families respectively.  The second
selection rule imposes independent constraints. It states that the
total $[s,t]$ degree has to be $[1,1]$. Of the two possible $[s,t]$
degrees associated with the Higgs fields, only the $[1,0]$ subspace
satisfies the $[s,t]$ selection rule. 
This selection rule disallows the second Higgs-Higgs conjugate pair
$(H_{2},\bar{H}_{2})$ from forming non-zero cubic Yukawa couplings.  
That is, the only non-vanishing $\lambda$ coefficients consistent with
both the $(p,q)$ and $[s,t]$ selection rules are of the form
\begin{equation}
  \lambda_{u,1j}^{1}, \lambda_{u,j1}^{1} \quad \lambda_{d,1j}^{1},
  \lambda_{d,j1}^{1} \quad 
  \lambda_{\nu,1j}^{1}, \lambda_{\nu,j1}^{1}
  \quad \lambda_{e,1j}^{1}, \lambda_{e,j1}^{1}
  \label{gamma}
\end{equation}
corresponding to the first Higgs pair $(H_{1},\bar{H}_{1})$.

As in~\cite{lssa}, let us analyze, for example, the Yukawa contribution to the up-quark
mass matrix. Assuming that $H_{1}$ gets a non-vanishing vacuum expectation
value $\langle H_{1} \rangle$ in its charge neutral component, this
contribution can be written as
\begin{equation}
  \left(
    \begin{array}{ccc}
      0 & \lambda_{u,12}^{1}\langle H_{1} \rangle & \lambda_{u,13}^{1}\langle H_{1} \rangle \\
      \lambda_{u,21}^{1}\langle H_{1} \rangle & 0 & 0 \\
      \lambda_{u,31}^{1}\langle H_{1} \rangle & 0 & 0 
    \end{array}
  \right)
  \label{final1}
\end{equation}
Using independent non-singular transformations on the $Q_{i}$ and
$u_{i}$ fields, one can find bases in which eq.~\eqref{final1} becomes
\begin{equation}
  \left(
    \begin{array}{ccc}
      0 &    0       &    0\\
      0 & \lambda\langle H_{1} \rangle &    0 \\
      0 &    0       & \lambda\langle H_{1} \rangle
    \end{array}
  \right)
  \label{final2}
\end{equation}
where $\lambda$ is an arbitrary, but non-zero, real number. We
conclude from the zero diagonal element that one up-quark is strictly
massless\footnote{At least, on the classical level. Higher order and
  non-perturbative terms in the superpotential could lead to naturally
  small corrections.}.  Furthermore, the two non-zero diagonal
elements imply that the second and third up-quarks will have
non-vanishing masses of $O\big(\langle H_{1} \rangle\big)$. However, the
exact value of their masses will depend on the explicit normalization
of the kinetic energy terms in the low-energy theory. These masses,
therefore, are in general not degenerate.  This analysis applies to
the down-quarks and the up- and down-leptons as well.  We conclude
that, prior to higher order and non-perturbative corrections, one
complete generation of quarks/leptons will be massless.  The remaining
two generations will have non-vanishing masses on the order of the
electroweak symmetry breaking scale which are, generically,
non-degenerate.

The coefficients $\lambda$ have no interpretation as an intersection
number and, therefore, no reason to be constant over the moduli space.
In general, we expect them to depend on the moduli. Of course, to
explicitly compute the quark/lepton masses one needs, in addition, the
Kahler potential, which determines the correct normalization of the
fields.


\section{Cubic $\mu$-Terms in the Superpotential}
\label{sec:couplings2}

In this section, we focus on computing Higgs--Higgs conjugate
$\mu$-terms.  First, note that in our heterotic model the two pairs of
Higgs fields arise from eq.~\eqref{120} as zero modes of the Dirac operator.
Hence, they cannot have ``bare'' $\mu$-terms in the superpotential
proportional to $H \bar{H}$ with constant coefficients. However,
$H$ and $\bar{H}$ can have cubic interactions
with the vector bundle moduli of the form $\phi H \bar{H}$.  If the
moduli develop non-vanishing vacuum expectation values, then these
cubic interactions generate $\mu$-terms of the form $\langle \phi
\rangle H \bar{H}$ in the superpotential.  Hence, we expect 
Higgs $\mu$-terms that are linearly dependent
on the vector bundle moduli. Classically, no higher dimensional
coupling of moduli to $H$ and $\bar{H}$ is allowed.

It follows from eq.~\eqref{120} that the $4$-dimensional Higgs and
moduli fields correspond to certain $\bar{\partial}$-closed
$(0,1)$-forms on $\Xt$ with values in the vector bundle $\wedge^2 \Vt$
and $\ad(\Vt)$ respectively. Denote these forms by $\Psi_H$,
$\Psi_{\bar{H}}$, and $\Psi_\phi$. They can be written as
\begin{equation}
  \Psi_H = \psi^{(H)}_{\bar{\iota}, [ab]} 
  \,\diff \bar{z}^{\bar{\iota}}
  , \qquad
  \Psi_{\bar{H}} = \psi^{(\bar{H})}_{\bar{\iota},[ab]}
  \,\diff \bar{z}^{\bar{\iota}}
  , \qquad
  \Psi_\phi = [\psi^{(\phi)}_{\bar{\iota}}]^{~b}_a
  \,\diff \bar{z}^{\bar{\iota}}
  , \qquad
\end{equation}
where $a$, $b$ are valued in the $SU(4)$ bundle $\Vt$ and
$\{z^\iota,\bar{z}^{\bar{\iota}}\}$ are coordinates on the Calabi-Yau
threefold $\Xt$. Doing the dimensional reduction of the
$10$-dimensional Lagrangian yields cubic terms in the superpotential
of the $4$-dimensional effective action. It turns out,
see~\cite{lssa}, that the coefficient of the cubic coupling
$\phi H \bar{H}$ is simply the unique way to obtain a number out of
the forms $\Psi_H$, $\Psi_{\bar{H}}$, and $\Psi_\phi$. That is,
\begin{equation}
  W = \cdots + \hat{\lambda} \phi H \bar{H}  
\end{equation}
where
\begin{equation}
  \label{eq:lambdaintegral2}
  \begin{split}
    \hat{\lambda} ~&= 
      \int_\Xt
      \Omega \wedge 
      \Tr\Big[ 
      \Psi_\phi \wedge \Psi_H \wedge \Psi_{\bar{H}} 
      \Big]
    = \\ &= 
      \int_\Xt
      \Omega \wedge 
      \Big(\epsilon^{acde} 
      [\psi^{(\phi)}_{\bar{\iota}}]^{~b}_a 
      \,
      \psi^{(H)}_{\bar{\kappa}, [bc]} 
      \,
      \psi^{(\bar{H})}_{\bar{\lambda},[de]}
      \Big)
      \diff \bar{z}^{\bar{\iota}} \wedge
      \diff \bar{z}^{\bar{\kappa}} \wedge
      \diff \bar{z}^{\bar{\lambda}}
  \end{split}
\end{equation}
and $\Omega$ is the holomorphic $(3,0)$-form. Similarly to the 
Yukawa couplings discussed above, we are
using the wedge product together with a contraction of the vector
bundle indices to obtain a product
\begin{multline}
  \label{eq:cup2}
  H^1\Big(\tx, \ad(\tv) \Big) \otimes 
  H^1\Big(\tx, \atv \Big) \otimes
  H^1\Big(\tx, \atv \Big) 
  \longrightarrow \\ \longrightarrow
  H^3\Big(\tx, \ad(\tv) \otimes \atv \otimes \atv \Big)
  \longrightarrow 
  H^3\Big(\tx, \cO_{\tx} \Big),  
\end{multline}
plus the fact that on the Calabi-Yau manifold $\Xt$
\begin{equation}
  H^3\Big(\Xt,\cO_{\Xt}\Big) =
  H^3\Big(\Xt, K_\Xt\Big) = 
  H_{\bar{\partial}}^{3,3}\Big(\Xt\Big) =
  H^6\Big(\Xt\Big)
\end{equation}
can be integrated over. If one were to use the heterotic string with
the ``standard embedding'', then the above product would simplify
further to the intersection of certain cycles in the Calabi-Yau
threefold. However, in our case there is no such description.

Hence, to compute $\mu$-terms we must first analyze the cohomology
groups
\begin{equation} 
  H^1\Big(\tx, \ad(\tv)\Big),~
  H^1\Big(\tx, \atv \Big),~
  H^3\Big(\tx, \cO_{\tx} \Big)
  \label{141}
\end{equation}
and the action of $\z3z3$ on these spaces. We then have to evaluate
the product in eq.~\eqref{eq:cup2}. As we will see in the following
sections, the two independent elliptic fibrations of $\Xt$ will force
most, but not all, products to vanish.

\subsection{The First Elliptic Fibration}
\label{sec:LSS1_2}

The Leray spectrial sequences for the first elliptic fibration $\tx
\stackrel{\pi_{2}}{\longrightarrow} B_2 $ was discussed in detail in
subsection \ref{sec:LSS1}.  Furthermore, the first Leray decomposition
for the sheaves $\cO_{\Xt}$ and $\wedge^{2} \Vt$ associated with the
volume form and Higgs fields were presented in eqns.~\eqref{200}
and~\eqref{eq:H1atv}, respectively.  To find the $\phi H \bar{H}$
cubic terms, one must additionally compute the first Leray
decomposition for the sheaf $\ad(\Vt)$ associated with the vector
bundle moduli.

\subsubsection{The First Leray Decomposition of the Moduli}

The (tangent space to the) moduli space of the vector bundle $\Vt$ is
$H^1(\tx, \ad(\tv))$. First, note
that $\ad(\tv)$ is defined to be the traceless part of $\tv \otimes
\tv^\ast$. But the trace is just the trivial line bundle, whose first
cohomology group vanishes. Therefore
\begin{equation}
  \label{eq:tensadtrace} 
  H^1\Big(\tx, \ad(\tv)\Big) = 
  H^1\Big(\tx, \Vt \otimes \Vt^\ast \Big) - 
  {\underbrace{H^1\Big(\tx, \cO_\tx \Big)}_{=0}} 
  .
\end{equation}
Since the action of the Wilson line on the ${\bf 1}$ representation of
$Spin(10)$ is trivial, one need only consider the $\z3z3$ invariant
subspace of this cohomology. That is, in the decomposition of the
index of the Dirac operator, eq.~\eqref{120}, the vector bundle moduli fields are
contained in
\begin{equation}
  \left( H^1\big(\tx, \ad(\tv) \big) \otimes
    \Rep{1} \right)^{\z3z3}   
  = 
  H^1\Big(\tx, \ad(\tv) \Big)^{\z3z3}   
  =
  H^1\Big(\tx, \tv\otimes\tv^\ast \Big)^{\z3z3}
  .
\end{equation}
In a previous paper~\cite{lss}, presented an explicity method for computing the
$(p,q)$ decomposition of $H^\ast(\Xt, \ad(\Vt))^{\ZZZ}$ 
from the complex of intertwined long exact sequences in 
which this cohomolgy is embedded. Here, we simply present the results
for our specific bundles with two Higgs pairs.


We find that the $H^1$ entries in the $\Xt\to B_2$ Leray tableau for
$H^{\ast}(\Xt, \tv \otimes \tv^{\ast})^{\z3z3}$ are
\begin{equation}
  \label{44} \vcenter{ \def\w{20mm} \def\h{6mm} \xymatrix@C=0mm@R=0mm{
      {\scriptstyle q=1} 
      & *=<\w,\h>[F]{ 9 } 
      & *=<\w,\h>[F]{ 4 } 
      & *=<\w,\h>[F]{ 0 } 
      \\ 
      {\scriptstyle q=0} 
      & *=<\w,\h>[F]{ 0 } 
      & *=<\w,\h>[F]{ 4 } 
      & *=<\w,\h>[F]{ 9 } 
      \\ 
      & {\scriptstyle p=0}
      & {\scriptstyle p=1} 
      & {\scriptstyle p=2} }} 
  \Rightarrow
  H^{p+q}\Big(\Xt, \tv \otimes \tv^{\ast}\Big)^{\z3z3}
\end{equation}
where, as previously, the non-zero entries denote the rank $4$ and $9$
trivial representations of \ZZZ. Note that
\begin{equation} 
  H^1(\Xt, \tv \otimes \tv^{\ast})^{\z3z3} 
  =
  9 + 4
  =
  13,
  \label{45}
\end{equation}
which is consistent with the statement in eq.~\eqref{moduli} that there
are a total of $13$ vector bundle moduli. Now, however, we have
determined the $(p,q)$ decomposition of $H^1(\Xt, \tv \otimes
\tv^{\ast})^{\z3z3}$ into the subspaces
\begin{equation} 
  \label{46}
  H^1\Big(\Xt,\tv \otimes \tv^{\ast}\Big)^{\z3z3}=
  \Hpq{\tv \otimes \tv^{\ast}}{0}{1}^{\z3z3}
  ~\oplus~
  \Hpq{\tv \otimes \tv^{\ast}}{1}{0}^{\z3z3}
  ,
\end{equation}
where
\begin{equation} 
  \Hpq{\tv \otimes \tv^{\ast}}{0}{1}^{\z3z3} = 9
  ,\quad
  \Hpq{\tv \otimes \tv^{\ast}}{1}{0}^{\z3z3} = 4
  ,
  \label{47}
\end{equation}
respectively.

\subsubsection{The (p,q) Selection Rule}

Having computed the decompositions of $H^3(\tx, \cO_{\tx} )$,
$H^1(\tx, \atv )$ and $H^1(\tx, \ad(\tv) )^{\z3z3}$ into their
${(p,q)}$ Leray subspaces, we can now analyze the $(p,q)$ components
of the triple product 
\begin{equation}
  \label{eq:product2}
  H^1\Big(\tx, \Vt \otimes \Vt^\ast \Big)^\z3z3 \otimes 
  H^1\Big(\tx, \atv \Big) \otimes
  H^1\Big(\tx, \atv \Big) 
  \longrightarrow 
  H^3\Big(\tx, \cO_{\tx} \Big)  
\end{equation}
given in eq.~\eqref{eq:cup2}. Inserting eqns.~\eqref{eq:H1atv}
and~\eqref{46}, we see that
\begin{multline}
  \label{eq:pqCup2}
  H^1\Big(\tx, \Vt \otimes \Vt^\ast \Big)^\z3z3 \otimes 
  H^1\Big(\tx, \atv \Big) \otimes
  H^1\Big(\tx, \atv \Big) 
  = \\ =
  \Big( \Hpq{\Vt \otimes \Vt^\ast}{0}{1} ^{\ZZZ} \oplus 
        \Hpq{\Vt \otimes \Vt^\ast}{1}{0} ^{\ZZZ} \Big)
  \otimes
  \Hpq{\atv}{1}{0} \otimes 
  \Hpq{\atv}{1}{0} 
  = \\ =
  \underbrace{\Big( 
    {\scriptstyle 
      \Hpq{\Vt \otimes \Vt^\ast}{0}{1}^{\z3z3} \otimes
      \Hpq{\atv}{1}{0} \otimes 
      \Hpq{\atv}{1}{0} 
    }
    \Big) 
  }_{\text{total $(p,q)$ degree }=\, (2,1)}
  \oplus
  \underbrace{\Big( 
    {\scriptstyle 
      \Hpq{\Vt \otimes \Vt^\ast}{1}{0}^\z3z3 \otimes
      \Hpq{\atv}{1}{0} \otimes 
      \Hpq{\atv}{1}{0} 
    }
    \Big) 
  }_{\text{total $(p,q)$ degree }=\, (3,0)}
  .
\end{multline}
Because of the $(p,q)$ degree, only the first term can have a
non-zero product in
\begin{equation}
  \label{480}
  H^3 \Big( \tx, \cO_\tx \Big) = \Hpq{\cO_\tx}{2}{1}
  ,
\end{equation}
see eq.~\eqref{200}. It follows that out of the $H^1(\tv \otimes
\tv^{\ast})^{\z3z3}=13$ vector bundle moduli, only
\begin{equation} 
  \Hpq{\tv \otimes \tv^{\ast}}{0}{1}^{\z3z3} = 9
  \label{490}
\end{equation}
will form non-vanishing Higgs--Higgs conjugate $\mu$-terms.  The
remaining $4$ moduli in the $\Hpq{\tv \otimes
  \tv^{\ast}}{1}{0}^{\z3z3}$ component have the wrong $(p,q)$ degree
to couple to a Higgs--Higgs conjugate pair.  As in the case of Yukawa
couplings, we refer to this as the $(p,q)$ Leray degree selection
rule. We conclude that the only non-zero product in
eq.~\eqref{eq:product2} is of the form
\begin{equation}
  \label{eq:pqproduct2}
  \Hpq{\Vt \otimes \Vt^\ast}{0}{1}^\z3z3 \otimes
  \Hpq{\atv}{1}{0} \otimes 
  \Hpq{\atv}{1}{0} 
  \longrightarrow
  \Hpq{\cO_\Xt}{2}{1}
  .
\end{equation}
Roughly what happens is the following. The Leray spectral sequence
decomposes differential forms into the number $p$ of legs in the
direction of the base and the number $q$ of legs in the fiber
direction. The holomorphic $(3,0)$-form $\Omega$ has two legs in the
base and one leg in the fiber direction. According to
eq.~\eqref{eq:H1atv}, both $1$-forms $\Psi_H$, $\Psi_{\bar{H}}$
corresponding to Higgs and Higgs conjugate have their one leg in the
base direction.  Therefore, the wedge product in
eq.~\eqref{eq:lambdaintegral2} can only be non-zero if the modulus
$1$-form $\Psi_\phi$ has its leg in the fiber direction, which only
$9$ out of the $13$ bundle moduli satisfy.

We conclude that due to a selection rule for the $(p,q)$ Leray degree,
the Higgs $\mu$-terms in the effective low-energy theory can involve
only $9$ of the $13$ vector bundle moduli.

\subsection{The Second Elliptic Fibration}
\label{sec:LSS2_2}

So far, we only made use of the fact that our Calabi-Yau manifold is
an elliptic fibration over the base $B_2$. But the $\dP9$ surface
$\B2$ is itself elliptically fibered over a $\IP1$. Consequently,
there is yet another selection rule coming from the second elliptic
fibration. The Leray spectral sequence for the second elliptic
fibration $\B2 \stackrel{\beta_{2}}{\longrightarrow} \IP1$ was
discussed in subsection \ref{sec:LSS2}.  Furthermore, the second Leray
decomposition for the sheaves $\cO_{\tx}$ and $\wedge^{2}\Vt$
associated with the volume form and Higgs fields were presented in
eqns.~\eqref{55} and~\eqref{beta}, respectively. To find the $\phi H
\Bar{H}$ cubic terms, one must additionally compute the second Leray
decomposition for the sheaf $\ad(\Vt)$ associated with the vector
bundle moduli.


\subsubsection{The Second Leray Decomposition of the Moduli}

Let us consider the $[s,t]$ Leray tableau for the moduli. We
have already seen that, due to the $(p,q)$ selection rule, only
\begin{equation}
  \Hpq{\tv \otimes \tv^{\ast}}{0}{1}^{\z3z3} = 
  9
  \quad \subset
  H^1\Big( \Xt, \Vt \otimes \Vt^\ast \Big)^\z3z3 
\end{equation}
out of the $13$ moduli can occur in the Higgs--Higgs conjugate
$\mu$-term. Therefore, we are only interested in the $[s,t]$
decomposition of this subspace; that is, the degree $0$ cohomology of
the sheaf $R^1\pi_{2\ast} \big( \Vt \otimes \Vt^\ast \big)$. The
corresponding Leray tableau is given by
\begin{equation}
  \label{610} 
  \vcenter{ \def\w{20mm} \def\h{6mm} \xymatrix@C=0mm@R=0mm{
      {\scriptstyle t=1} & 
      *=<\w,\h>[F]{  } & 
      *=<\w,\h>[F]{  } \\
      {\scriptstyle t=0} & 
      *=<\w,\h>[F]{ 9 } & 
      *=<\w,\h>[F]{  }
      \\ & {\scriptstyle s=0} & {\scriptstyle s=1} }} 
  \Rightarrow
  H^{s+t}\Big(B_2, R^1\pi_{2\ast} \big(\Vt\otimes\Vt^\ast\big)
  \Big)^\z3z3     
  ,
\end{equation}
where the empty boxes are of no interest for our purposes. It follows
that the $9$ moduli of interest have $[s,t]$ degree $[0,0]$. That is,
\begin{equation} 
  \label{620}
  \Hpq{\tv \otimes \tv^{\ast}}{0}{1}^{\z3z3} = 
  \Hst{\tv \otimes \tv^{\ast}}{1}{0}{0}^{\z3z3} = 
  9
  .
\end{equation}

\subsubsection{The [s,t] Selection Rule}

Having computed the decompositions of the relevant cohomology spaces
into their $[s,t]$ Leray subspaces, we can now calculate the triple
product eq.~\eqref{eq:cup2}. The $(p,q)$ selection rule dictates that
the only non-zero product is of the form eq.~\eqref{eq:pqproduct2}. Now
split each term in this product into its $[s,t]$ subspaces, as given
in eqns.~\eqref{55},~\eqref{beta}, and~\eqref{620} respectively. The
result is
\begin{multline} 
  \label{650}
  \Hst{\tv \otimes \tv^{\ast}}{1}{0}{0}^{\z3z3} 
  \otimes
  \Big( \Hst{\atv}{0}{0}{1} \oplus \Hst{\atv}{0}{1}{0} \Big)
  \otimes \\ \otimes
  \Big( \Hst{\atv}{0}{0}{1} \oplus \Hst{\atv}{0}{1}{0} \Big)
  \longrightarrow
  \Hst{\cO_\Xt}{1}{1}{1} 
  .
\end{multline}
Clearly, this triple product vanishes by degree unless we choose the
$\Hst{\atv}{0}{0}{1}$ from one of the $\Hpq{\atv}{1}{0}$ subspaces and
$\Hst{\atv}{0}{1}{0}$ from the other.  In this case, eq.~\eqref{650}
becomes
\begin{equation} 
  \label{650-1}
  \Hst{\tv \otimes \tv^{\ast}}{1}{0}{0}^{\z3z3} 
  \otimes
  \Hst{\atv}{0}{1}{0}
  \otimes
  \Hst{\atv}{0}{0}{1}
  \longrightarrow
  \Hst{\cO_\Xt}{1}{1}{1} 
  ,
\end{equation}
which is consistent.

\subsubsection{Wilson Lines}

Recall that we have, in addition to the $SU(4)$ instanton, a
Wilson line\footnote{In fact, we switch on a separate Wilson line for
  both $\Z_3$ factors in $\pi_1(X)=\z3z3$.} turned on. Its effect is
to break the $Spin(10)$ gauge group down to the desired $SU(3)_{C}
\times SU(2)_{L} \times U(1)_{Y} \times U(1)_{B-L}$ gauge group. Each
fundamental matter field in the $\Rep{10}$ can be broken to a Higgs
field, a color triplet, or projected out. The $\ZZZ$ action of
the Wilson line $W$ on a $\Rep{10}$ representation of $Spin(10)$
was given in \eqref{66}.  Tensoring this with
the cohomology space $H^1\big(\Xt,\atv\big)$ presented
in~\eqref{beta},\eqref{58}, we found the invariant
subspace under the combined $\z3z3$ action on the cohomology space and
the Wilson line to be
\begin{equation}
  \Big( H^1\big(\Xt,\atv\big) \otimes \Rep{10} \Big)^\z3z3
  = 
  \Span
  \big\{ H_1, \bar{H}_1, H_2, \bar{H}_2 \big\}
  .
\end{equation}
That is, two copies of Higgs and two copies of Higgs
conjugate fields survive the $\z3z3$ quotient. As required for any
realistic model, all color triplets are projected out.

Further information was obtained from the $(p,q)$ and $[s,t]$ degrees of the
Higgs fields. Using the decomposition of $H^1\big(\Xt,\atv\big)$, we
found
\begin{multline}
  \label{decomp}
  \Big( H^1\big(\Xt,\atv\big) \otimes \Rep{10} \Big)^\z3z3
  =
  \Big( \Hpq{\atv}{1}{0} \otimes \Rep{10} \Big)^\z3z3
  = \\ =
  \underbrace{
    \Big( \Hst{\atv}{0}{0}{1} \otimes \Rep{10} \Big)^\z3z3
  }_{= \Span\{H_2, \bar{H}_2\} }
  \oplus 
  \underbrace{
    \Big( \Hst{\atv}{0}{1}{0} \otimes \Rep{10} \Big)^\z3z3
  }_{= \Span\{H_1, \bar{H}_1\} }
  .
\end{multline}
Recall that the $(H_1,\Bar{H}_1)$ Higgs pair can form non-vanishing 
cubic Yukawa couplings, whereas the $(H_2,\bar{H}_2)$ pair is
forbidden to do so by the $[s,t]$ selection rule.  

Decomposition~\eqref{decomp} also labels the cubic $\mu$-term coupling
to the moduli.  Note that $[s,t]$ selection rule eq.~\eqref{650-1}
only allows non-vanishing cubic $\mu$-terms involving one Higgs field
from $\Hst{\atv}{0}{0}{1}$ and one Higgs field from
$\Hst{\atv}{0}{1}{0}$. It follows that the cubic $\mu$-terms are of
the form $\phi H_{1}\bar{H}_{2}$ and $\phi\bar{H}_{1} H_{2}$ only.

\subsection{\texorpdfstring{Higgs $\mu$-terms}{Higgs mu-terms}}

To conclude, we analyzed cubic terms in the superpotential of the form
\begin{equation} 
  \hat{\lambda}^{m}_{kl} \phi_{m} H_{k} \bar{H}_{l}
  ,
  \label{670-1}
\end{equation}
where
\begin{itemize}
\item $\hat{\lambda}^{m}_{kl}$ is a coefficient determined by the integral
  eq.~\eqref{eq:lambdaintegral2},
\item $\phi_m,~ m=1,\dots,13$ are the vector bundle moduli,
\item $H_k,~ k=1,2$ are the two Higgs fields, and
\item $\bar{H}_l,~  l=1,2$ are the two Higgs conjugate fields.
\end{itemize}
We found that they are subject to two independent selection rules
coming from the two independent torus fibrations. The first selection
rule is that the total $(p,q)$ degree is $(2,1)$. According to
~\eqref{eq:pqproduct2}, $H_k\bar{H}_l$ already has $(p,q)$ degree
$(2,0)$. Hence the moduli fields $\phi_m$ must have degree $(0,1)$. In
eq.~\eqref{47} we found that only the moduli $\phi_m$,
$m=1,\dots,9$, have the right $(p,q)$ degree. In other words,
the coefficients
\begin{equation}
  \label{lambda2}
  \hat{\lambda}^{m}_{kl}=0
  , 
  \qquad m=10, \dots, 13
\end{equation}
must vanish. Furthermore, the second selection rule eq.~\eqref{650-1}
imposes independent constraints. It states that the total $[s,t]$
degree has to be $[1,1]$. We showed that only the cubic terms
$\phi_{m} H_{1} \bar{H}_{2}$ and $\phi_{m}\bar{H}_1 H_2$ for
$m=1,\dots,9$.  have the correct degree $[1,1]$.  Therefore,
the $(p,q)$ and $[s,t]$ selection rules allow $\mu$-terms involving
$9$ out of the $13$ vector bundle moduli coupling to
$H_{1}\bar{H}_{2}$ and $H_{2}\bar{H}_{1}$, but disallow their coupling
to $H_{1}\bar{H}_{1}$ and $H_{2}\bar{H}_{2}$. Cubic terms involving
Higgs--Higgs conjugate fields with any of the remaining
$4$ moduli are forbidden in the superpotential.  That
is, the only non-vanishing $\hat{\lambda}$ coefficients in
\eqref{670-1} are of the form
\begin{equation}
  \hat{\lambda}^{m}_{12},\hat{\lambda}^{m}_{21},\quad m=1,\dots,9.
\label{add1}
\end{equation}
Note that the expressions~\eqref{lambda2} and~\eqref{add1} naturally partition the $m=1,\dots,13$ index into
\begin{equation}
\{m\}=\{{\bar{m}}, {\tilde{m}}\} \; ,
\label{add2}
\end{equation}
where ${\bar{m}}=1,\dots,9$ and ${\tilde{m}}=10,\dots,13$. 
When the moduli develop non-zero vacuum expectation values, these
superpotential terms generate Higgs $\mu$-terms of the form
\begin{equation} 
  \hat{\lambda}^{\bar{m}}_{12} \, \langle \phi_{\bar{m}}\rangle \, 
  H_{1} \bar{H}_{2} +
  \hat{\lambda}^{\bar{m}}_{21} \, \langle \phi_{\bar{m}}\rangle \, 
  H_{2} \bar{H}_{1}
  ,
  \qquad 
  \bar{m}=1,\dots, 9
  .
  \label{67-2}
\end{equation}

The coefficients $\hat{\lambda}^{m}_{kl}$ have no interpretation as
intersection numbers and, therefore, no reason to be constant over
moduli space. In general, we expect them to depend on the moduli. Of
course, to explicitly compute these functions one needs the Kahler
potential which determines the correct normalization for all fields.

\section{Discussion of the Superpotential}

As shown in the previous two sections, the perturbative holomorphic 
superpotential for zero-modes of the two Higgs-Higgs conjugate pair vacua presented 
in this paper is given, up to operators of dimension $4$, by
\begin{equation}
  W_{0}=W_\text{Yukawa}+W_{\mu},
\label{burt1}
\end{equation}
where 
\begin{equation}
  W_\text{Yukawa}= 
  {\lambda}^{1}_{u,ij}Q_{i}{H}_{1}u_{j}+
  {\lambda}^{1}_{d,ij}Q_{i}\bar{H}_{1}d_{j}+
  {\lambda}^{1}_{\nu,ij}L_{i}{H}_{1}\nu_{j}+
  {\lambda}^{1}_{e,ij}L_{i}\bar{H}_{1}e_{j}
\label{burt2}
\end{equation}
with the restriction $i=1,j=2,3$ or $i=2,3,j=1$, and
\begin{equation}
W_{\mu}=  {\hat{\lambda}}^{\bar{m}}_{12}\phi_{\bar{m}} {H}_{1}\bar{H}_{2}+ 
         {\hat{\lambda}}^{\bar{m}}_{21}\phi_{\bar{m}} H_{2}\bar{H}_{1},
\label{burt3}
\end{equation}
where $\bar{m}=1,\dots,9$.  Quadratic mass terms do not appear in $W_{0}$
since all fields in the perturbative low energy theory are strictly
zero-modes of the Dirac operator.  Furthermore, the cubic terms are
restricted by the ``stringy'' $(p,q)$ and $[s,t]$ Leray selection
rules. Specifically, non-vanishing Yukawa terms can only occur between
the first family of quarks/leptons and the second and third
quark/lepton families. In addition, only the first pair of Higgs-Higgs
conjugate fields, $H_{1}$ and $\bar{H}_{1}$, can appear in these
non-vanishing Yukawa couplings. Similary, non-zero cubic $\mu$-terms
can only occur beween a specific $9$-dimensional subset of the $13$
vector bundle moduli and the restricted pairs $H_{1}\bar{H}_{2}$ and
$H_{2}\bar{H}_{1}$.

It is important to note, however, that only the zero-modes need have 
vanishing mass terms. Non zero-modes, that is, the superfields 
corresponding to Kaluza-Klein states, do add quadratic terms to the 
superpotential. For example, let ${\bf H}$ and 
$\bar{\bf H}$ be two superfields corresponding to Kaluza-Klein modes
with the same quantum numbers as $H_{1,2}$ and $\bar{H}_{1,2}$. These contribute
a mass term
\begin{equation}
W_{\text{mass}, KK}=M_{c} {\bf H} \bar{\bf H}
\label{burt4}
\end{equation}
to the superpotential, where $M_{c}$ is of the order of the Calabi-Yau
compactification scale. Similarly, the $(p,q)$ and $[s,t]$ Leray
selection rules only apply to the cubic product of the sheaf
cohomologies associated with the zero-modes of the Dirac operator.  It
follows that there is no restraint, other than group theory, on cubic
terms involving at least one Kaluza-Klein superfield. The terms of
interest for this paper are
\begin{equation}
W_{\text{Yukawa}, KK}= \tilde{\lambda}_{u,ij}Q_{i}{\bf H}u_{j}+
        \tilde{\lambda}_{d,ij}Q_{i}\bar{\bf H}d_{j}+
        \tilde{\lambda}_{\nu,ij}L_{i}{\bf H}\nu_{j}+
        \tilde{\lambda}_{e,ij}L_{i}\bar{\bf H}e_{j}
\label{burt5}
\end{equation}
and
\begin{equation}
W_{\mu,KK}= \tilde{\hat{\lambda}}^{m}_{k}\phi_{m} {\bf H}\bar{H}_{k}+ 
          \tilde{\hat{\lambda}}^{'m}_{k}\phi_{m} H_{k}\bar{\bf H},
\label{burt6}
\end{equation}
where the sums over $i,j=1,2,3$ as well as $m=1,\dots,13$ and $k=1,2$
are unconstrained.

The significance of this is that such interactions can quantum mechanically
induce amplitudes which, at energy small compared to the compactification 
scale, appear as irreducible, holomorphic higher-dimensional contributions to
the superpotential. Despite the fact that such terms depend on zero-modes 
only, they are not subject to $(p,q)$ and $[s,t]$ selection rules
since they are not generated as a triple cohomology product. 
There are two classes of tree-level supergraphs that are of particular interest for this paper.
The first of these is shown in Figure 1.
An analysis of these graphs shows that for energy-momenta much less 
than the compactification scale, that is, $k^{2} \ll M_{c}^{2}$, they induce
quartic terms in the superpotential of the form

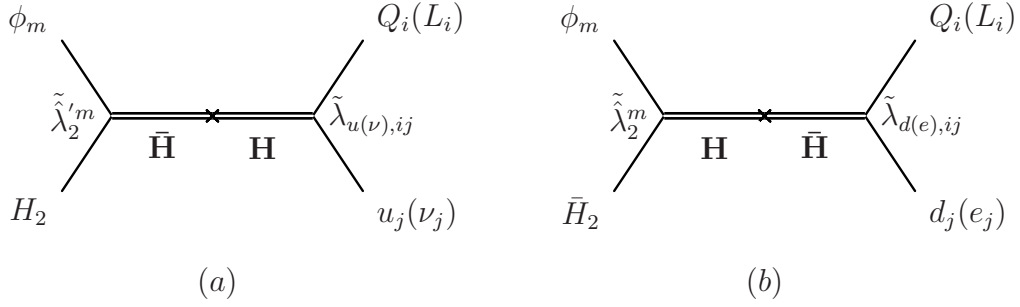
\begin{figure}
  \centering
\begin{fmffile}{KKmode}
  \newenvironment{treegraph}
  {\begin{fmfgraph*}(50,20)
    \fmfleft{i1,i2}
    \fmfright{o1,o2}
    \fmf{vanilla}{i1,v1,i2}
    \fmf{vanilla}{o1,v3,o2}
    \fmfv{decor.shape=cross,decor.size=3thick}{v2}  }
  {\end{fmfgraph*}}
$\begin{array}{c c c }

 \fmfframe(0,0)(0,0){
  \begin{treegraph}
    \fmflabel{$\phi_{m}$}{i2}
    \fmflabel{$H_{2}$}{i1}
    \fmflabel{$Q_{i}(L_{i})$}{o2}
    \fmflabel{$u_{j}(\nu_{j})$}{o1}
    \fmflabel{$\tilde{\hat{\lambda}}^{'m}_{2}$}{v1}
    \fmflabel{$\tilde{\lambda}_{u(\nu),ij}$}{v3}
    \fmf{double,label=${\bf \bar{H}}$}{v1,v2}
    \fmf{double,label=${\bf H}$,label.dist=8}{v2,v3}
  \end{treegraph} 
 } 
&
    \hspace{15mm}
& 
  \begin{treegraph}
    \fmflabel{$\phi_{m}$}{i2}
    \fmflabel{$\bar{H}_{2}$}{i1}
    \fmflabel{$Q_{i}(L_{i})$}{o2}
    \fmflabel{$d_{j}(e_{j})$}{o1}
    \fmflabel{$\tilde{\hat{\lambda}}^{m}_{2}$}{v1}
    \fmflabel{$\tilde{\lambda}_{d(e),ij}$}{v3}

    \fmf{double,label=${\bf H}$,label.dist=8}{v1,v2}
    \fmf{double,label=${\bf \bar{H}}$}{v2,v3}
  \end{treegraph}
\\
\\
(a) &   & (b) 
\end{array}$
\end{fmffile}
   \caption{Kaluza-Klein mode mediated supergraphs giving rise to $W_{4}$
and effective Yukawa couplings of quarks/leptons to the second Higgs pair.}
  \label{fig:KKmodes}
\end{figure}

\begin{equation}
W_{4}= \tilde{\lambda}_{u,ij} \tilde{\hat{\lambda}}^{'m}_{2} 
\frac{\phi_{m}}{M_{c}}Q_{i}H_{2}u_{j}+
\tilde{\lambda}_{d,ij}  \tilde{\hat{\lambda}}^{m}_{2} 
\frac{\phi_{m}}{M_{c}}Q_{i}\bar{H}_{2}d_{j}+
\tilde{\lambda}_{\nu,ij} \tilde{\hat{\lambda}}^{'m}_{2}
\frac{\phi_{m}}{M_{c}}L_{i}H_{2}\nu_{j}+ 
 \tilde{\lambda}_{e,ij} \tilde{\hat{\lambda}}^{m}_{2}
\frac{\phi_{m}}{M_{c}}L_{i}\bar{H}_{2}e_{j},
\label{burt7}
\end{equation}
where the sums over $m=1,\dots,13$ and $i,j=1,2,3$ are
unrestricted. These terms are of physical significance since, if at
least one of the vector bundle moduli has a non-vanishing vacuum
expectation value $\langle\phi_{m}\rangle$, they yield cubic Yukawa
terms where quark/lepton superfields couple to the second Higgs pair,
$H_{2}$ and $\bar{H}_{2}$. The induced Yukawa interactions are of the
form
\begin{equation}
W_{4, \text{Yukawa}}={\lambda}^{2}_{u,ij}Q_{i}H_{2}u_{j}+
 {\lambda}^{2}_{d,ij}Q_{i}\bar{H}_{2}d_{j}+
 {\lambda}^{2}_{\nu,ij}L_{i}H_{2}\nu_{j}+
 {\lambda}^{2}_{e,ij}L_{i}\bar{H}_{2}e_{j},
\label{burt8}
\end{equation}
where
\begin{equation}
{\lambda}^{2}_{u(\nu),ij}=\tilde{\lambda}_{u(\nu),ij} \tilde{\hat{\lambda}}^{'m}_{2} \frac{\langle\phi_{m}\rangle}{M_{c}}, \quad 
{\lambda}^{2}_{d(e),ij}= \tilde{\lambda}_{d(e),ij} \tilde{\hat{\lambda}}^{m}_{2}  \frac{\langle\phi_{m}\rangle}{M_{c}}
\; .
\label{burt8a}
\end{equation}
Such couplings were disallowed classically by the $(p,q)$ and $[s,t]$ Leray
selection rules, as discussed above, but can be generated from the quartic
terms in $W_{4}$ when the vector bundle moduli have non-vanishing expectation values.
It is important to note, however, that since these Yukawa couplings to the second Higgs pair 
arise from higher dimension operators, they are naturally suppressed by the factors
\begin{equation}
\tilde{\hat{\lambda}}^{'m}_{2} \frac{\langle\phi_{m}\rangle}{M_{c}} \ll 1 \; , \quad 
\tilde{\hat{\lambda}}^{m}_{2}  \frac{\langle\phi_{m}\rangle}{M_{c}} \ll1 \; .
\label{home1}
\end{equation}
An estimate of the magnitudes of these factors will be presented below. Let us assume, for example, that 
the cubic couplings of quarks/leptons to the Kaluza-Klein Higgs pair ${\bf H},{\bf{\bar{H}}}$ 
are of the same order of magnitude as their Yukawa couplings to $H_{1},{\bar{H}}_{1}$; that is, $\tilde{\lambda}_{u(\nu),ij} \sim  \lambda^{1}_{u(\nu),ij}$, $\tilde{\lambda}_{d(e),ij} \sim \lambda^{1}_{d(e),ij}$. 
Then it follows from~\eqref{home1} that
\begin{equation}
\lambda^{2}_{u(\nu),ij}  \ll \lambda^{1}_{u(\nu),ij} \;, \quad \lambda^{2}_{d(e),ij} \ll \lambda^{1}_{d(e),ij} \; . 
\label{home2}
\end{equation}
Clearly this will remain true for a much wider range of assumptions as well, depending on the magnitude of the suppression factors in~\eqref{home1}. We conclude that the Yukawa couplings of quarks/leptons to the second Higgs pair are {\it naturally suppressed} relative to the 
Yukawa couplings to the first Higgs pair. The physical implications of this will be discussed in detail below. Before doing that, however,  let us provide an estimate for the suppression factors in~\eqref{home1}.

The second class of supergraphs of interest is shown in Figure 2. In 
the low energy-momentum limit, $k^{2} \ll M_{c}^{2}$, these induce quartic terms in the superpotential of the form
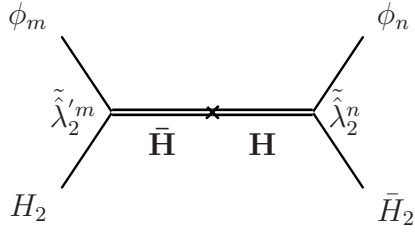
\begin{figure}
\centering
\begin{fmffile}{modulimodes}
  \newenvironment{treegraph}
  {\begin{fmfgraph*}(50,20)
    \fmfleft{i1,i2}
    \fmfright{o1,o2}
    \fmf{vanilla}{i1,v1,i2}
    \fmf{vanilla}{o1,v3,o2}
    \fmfv{decor.shape=cross,decor.size=3thick}{v2}  }
  {\end{fmfgraph*}}
\fmfframe(0,20)(0,0){
    \begin{treegraph}
    \fmflabel{$\phi_{m}$}{i2}
    \fmflabel{$H_{2}$}{i1}
    \fmflabel{$\phi_{n}$}{o2}
    \fmflabel{$\bar{H}_{2}$}{o1}
    \fmflabel{$\tilde{\hat{\lambda}}^{'m}_{2}$}{v1}
    \fmflabel{$\tilde{\hat{\lambda}}^{n}_{2}$}{v3}
    \fmf{double,label=${\bf \bar{H}}$}{v1,v2}
    \fmf{double,label=${\bf H}$,label.dist=8}{v2,v3}
  \end{treegraph}    
}
\end{fmffile}
   \caption{Kaluza-Klein mode mediated supergraphs giving rise to $W'_{4}$
and effective $\mu$ terms in the superpotential. }
  \label{fig:KKphimodes}
\end{figure}

\begin{equation}
W'_{4}=\tilde{\hat{\lambda}}^{'m}_{k}\tilde{\hat{\lambda}}^{n}_{l}\frac{\phi_{m}}{M_{c}}\phi_{n} H_{k}{\bar{H}}_{l} \; ,
\label{sailing1}
\end{equation}
where the sums over $m,n=1,\dots,13$ and $k,l=1,2$ are unrestricted. These terms are physically significant since, if at least one of the vector bundle moduli has a non-vanishing vacuum expectation value $\langle \phi_{m} \rangle$, they induce Higgs $\mu$-terms of the form
\begin{equation}
W_{4,\mu}=\mu_{kl}H_{k}{\bar{H}}_{l} 
\label{sailing2}
\end{equation}
with coefficients
\begin{equation}
\mu_{kl}=\big(\tilde{\hat{\lambda}}^{'m}_{k}\frac{\langle\phi_{m}\rangle}{M_{c}}\big)\big(\tilde{\hat{\lambda}}^{n}_{l}\frac{\langle\phi_{n}\rangle}{M_{c}}\big) M_{c} \; .
\label{sailing3}
\end{equation}
On generic grounds, if this theory is to naturally have appropriate electroweak symmetry breaking, these $\mu$-coefficients must satisfy
\begin{equation}
\mu_{kl} \lesssim M_{EW} \; ,
\label{home3}
\end{equation}
where $M_{EW} \approx 10^{2}GeV$. It follows from~\eqref{sailing3} that
\begin{equation}
\tilde{\hat{\lambda}}^{'m}_{k}\frac{\langle\phi_{m}\rangle}{M_{c}} \sim \tilde{\hat{\lambda}}^{m}_{k}\frac{\langle\phi_{m}\rangle}{M_{c}} \lesssim \sqrt{\frac{M_{EW}}{M_{c}}} \approx 10^{-7} \; .
\label{home4}
\end{equation}
In the final term, we have chosen $M_{c} \approx 10^{16} GeV$. This is consistent with the inequalities~\eqref{home1} and gives a natural estimate for their magnitude. Note that if this bound is saturated, the natural suppression~\eqref{home2} of the Yukawa couplings to the second Higgs pair will remain true even if the $\tilde{\lambda}_{u(\nu),ij}$,
$\tilde{\lambda}_{d(e),ij}$ coupling parameters in~\eqref{burt8a}
 are as large as $\tilde{\lambda}_{u(\nu),ij} \sim \tilde{\lambda}_{d(e),ij} \sim 1$. In this case, one would have
\begin{equation}
\lambda^{2}_{u(\nu),ij}  \sim 10^{-7}\;, \quad \lambda^{2}_{d(e),ij} \sim 10^{-7} \; ,
\label{home4a}
\end{equation}
a fact we will use in the next section.

Let us now return to the low-energy theory described strictly by the zero-modes of the Dirac
operator. The Kaluza-Klein superfields ``decouple'' and, hence, we can ignore all 
interactions containing at least one of these heavy fields. It follows that the relevant superpotential for the low-energy theory is given by
\begin{equation}
W=W_\text{Yukawa}+W_{\mu}+W_{4}+W'_{4},
\label{burt8c}
\end{equation}
where $W_\text{Yukawa}$, $W_{\mu}$, $W_{4}$ and $W'_{4}$ are given in eqns.~\eqref{burt2},~\eqref{burt3},~\eqref{burt7} and~\eqref{sailing1} respectively.
In broad outline, the physics described by the superpotential $W$ in
\eqref{burt8c}, relevant to the fact that there are {\it two}
Higgs-Higgs conjugate pairs, is the following. First, note that since
the coefficients of the Yukawa couplings to the second Higgs pair,
$H_{2}$ and $\bar{H}_{2}$, are suppressed, it follows that the masses
of quarks and leptons are predominantly generated by the vacuum
expectation values of the first Higgs pair, $H_{1}$ and $\bar{H}_{1}$,
as in the standard MSSM. Second, the masses of the $W^{\pm}$ and $Z$
vector bosons receive contributions from both pairs of Higgs-Higgs
conjugate superfields through their respective kinetic energy
terms. Despite this, the GIM mechanism continues to apply at tree
level and, hence, $Z$ couples only to flavor {\it preserving}
currents.  Third, recall that in the single Higgs pair MSSM, all
flavor-changing currents coupled to the neutral Higgs scalar boson
vanish. This is no longer true, however, when the spectrum contains a
second Higgs pair. In this case, one expects Higgs-induced flavor
changing neutral currents coupled to as many as three neutral Higgs
bosons. If the coefficients of the Yukawa couplings to $H_{2}$ and
$\bar{H}_{2}$ were arbitrarily large, then these Higgs-induced neutral
currents would violate current phenomenological bounds on a number of
processes. However, the coefficients in $W_{4,Yukawa}$ in~\eqref{burt8} are {\it not}
arbitrarily large.  Rather, as mentioned above, they are all naturally suppressed by the factors presented in~\eqref{home1} and estimated in~\eqref{home4}.
Hence, if these factors are sufficiently
small the Higgs-induced flavor-changing neutral currents will be
consistent with present experimental data. Be that as it may,
they may still be sufficiently large in some region of parameter space to become
relevant as the precision of this data is improved.

A complete analysis of these issues would require the computation of the perturbative Kahler 
potential, the non-perturbative contributions to both the Kahler potential and the
superpotential, stabilization of all moduli, a complete exposition of supersymmetry
breaking and the explicit computation of electroweak and
$U(1)_{B-L}$ symmetry breaking. Although much of the theory required to accomplish this
already exists, it is clearly a long term project that we will not begin to attempt 
in this paper. Rather, we will explore the relevant physics within the context of a 
toy model which contains most of the salient features of our two Higgs pair vacua. To make this toy model as simple as possible, we close this section by noting from $W_{\mu}$ in~\eqref{burt3} that any non-vanishing vacuum expectation values $\langle \phi_{\bar{m}} \rangle$, $\bar{m}=1,\dots,9$ will induce $\mu$-terms of the form
\begin{equation}
W_{\mu}=\mu_{12}{H}_{1}\bar{H}_{2}+ \mu_{21}H_{2}\bar{H}_{1}+\dots,
\label{home5}
\end{equation}
where
\begin{equation}
\mu_{12}={\hat{\lambda}}^{\bar{m}}_{12}\langle\phi_{\bar{m}}\rangle \; , \quad 
 \mu_{21}={\hat{\lambda}}^{\bar{m}}_{21}\langle\phi_{\bar{m}}\rangle \; .
 \label{home6}
 \end{equation}
Exactly as in~\eqref{home3}, these $\mu$-coefficients must satisfy
\begin{equation}
\mu_{12}\; , \; \mu _{21} \lesssim M_{EW} \; 
\label{home7}
\end{equation}
and, hence,
\begin{equation}
{\hat{\lambda}}^{\bar{m}}_{12}\frac{\langle\phi_{\bar{m}}\rangle}{M_{c}} \sim {\hat{\lambda}}^{\bar{m}}_{21}\frac{\langle\phi_{\bar{m}}\rangle}{M_{c}}  \lesssim \frac{M_{EW}}{M_{c}} \approx 10^{-14} \;.
\label{home8}
\end{equation}
Assuming the parameters ${\hat{\lambda}}^{\bar{m}}_{12}$ and ${\hat{\lambda}}^{\bar{m}}_{21}$ are of order unity, or, at least, not extremely small, it follows from~\eqref{home4} that the contribution of the first $\bar{m}=1,\dots,9$ moduli to the induced Yukawa couplings ${\lambda}^{2}_{u(\nu),ij}$ and
${\lambda}^{2}_{d(e),ij}$ in~\eqref{burt8a} can be ignored. Since in this remainder of this paper we are concerned only with possible Higgs-mediated flavor-changing neutral currents, it is reasonable to simply drop all terms in the superpotential~\eqref{burt8c} containing these nine moduli and only consider terms with the four moduli $\phi_{\tilde{m}}$ with $\tilde{m}=10,\dots,13$. When constructing the toy model in the next section, we will base it on this truncated supersymmetric theory.

\section{A Simplified Model}

Much of the technical difficulty in analyzing our two Higgs pair string
vacua comes from the $N=1$ local supersymmetry. Great simplification is 
achieved, while retaining the relevant physics, by choosing our toy
model to be non-supersymmetric. We will also, for simplicity, ignore the 
$U(1)_{B-L}$ gauge symmetry, since its inclusion would not alter our 
conclusions. That is, we take our gauge group to be 
the $SU(3)_{C} \times SU(2)_{L}\times U(1)_{Y}$ of the standard model. 
Hence, after electroweak symmetry breaking our vector boson spectrum consists 
of three massive bosons, $W^{\pm}$,$Z$ and the massless photon $A$.

\subsection{The Spectrum}

We begin by including all of the matter fields of the standard model. 
That is, the spectrum contains three families of quark and lepton fermions,
each family transforming as
\begin{equation}
  \ Q = \big(\Rep{3},   \Rep{2}, 1 \big) \,,\quad
  \ u =\big(\Rep{3},\Rep{1}, 4 \big) \,,\quad
  \ d =\big(\Rep{3},\Rep{1}, -2 \big)
  \label{volker1}
\end{equation}
and
\begin{equation}
  \ L=\big(\Rep{1},\Rep{2}, -3 \big) \,,\quad
  \ e=\big(\Rep{1},\Rep{1}, -6 \big) \,,\quad
  \nu=\big(\Rep{1},\Rep{1}, 0 \big)
  \label{volker2}
\end{equation}
under $SU(3)_{C} \times SU(2)_{L} \times U(1)_{Y}$.
We have displayed the quantum number $3Y$ for
convenience. Note from eqn.~\eqref{volker2} that each family contains a
right-handed neutrino.

To complete the standard model spectrum, we add a complex Higgs scalar boson
which transforms as
\begin{equation}
H_{1}=\big( {\bf 1}, {\bf 2}, 3 \big)
\label{volker3}
\end{equation}
under the gauge group. This naturally forms Yukawa terms with the 
``up'' quark and neutrino singlets, whereas the ``down'' quark and 
lepton singlets couple to $H_{1}^{\ast}$ This is unlike the 
supersymmetric case, where one must introduce an independent $\bar{H}_{1}$
superfield.

So far, our toy model is exactly the standard model. However, to reflect the 
physics of our two Higgs pair string vacua, we now make several important
additions to the spectrum. First, in analogy with the 
second Higgs-Higgs conjugate pair $H_{2}$,$\bar{H}_{2}$, 
we introduce a second complex 
Higgs boson field $H_{2}$ (and, hence, $H_{2}^{\ast}$), transforming as
\begin{equation}
H_{2}=\big( {\bf 1}, {\bf 2}, 3 \big).
\label{volker4}
\end{equation}
Second, to play the role of the vector bundle moduli in the string vacua, we must add 
gauge singlet scalar fields to the spectrum. Recall that there are thirteen such moduli fields, which break into two types; nine that are allowed by the $(p,q)$ and 
$[s,t]$ selection rules to form cubic $\mu$-terms with the Higgs fields 
and four that are not. As discussed above, the moduli that form cubic $\mu$-terms give a sub-dominant contribution to the Yukawa couplings to the second Higgs pair and, for the purposes of this paper, can be ignored. Hence, we will not introduce them into our toy model. On the other hand, those moduli that are disallowed from forming cubic $\mu$-terms give the dominant contribution to these Yukawa couplings and must be part of the analysis. Therefore, we include them in the toy model.
For simplicity, we add a single, real scalar field $\phi$ to the spectrum to represent this 
type of field. As do moduli, this transforms
trivially as
\begin{equation}
\phi =\big( {\bf 1}, {\bf 1}, 0 \big)
\label{volker5}
\end{equation}
under the gauge group. Choosing this field to be complex and/or adding 
more than one such field would greatly complicate the analysis without 
altering the conclusion. 

\subsection{Discrete Symmetry}

If this model had no further restrictions, one would 
generically find, after electroweak symmetry breaking, flavor 
changing currents coupling with large coefficients to the neutral 
Higgs bosons. These Higgs mediated flavor-changing neutral currents would
easily violate the experimental bounds on a large number of physical
processes. As shown long ago~\cite{weinberg}, this problem can be naturally resolved
in two ways. First, one can introduce a discrete symmetry which only allows
Yukawa couplings of ``up'' quark and neutrino singlets to $H_{1}$ and
``down'' quark and lepton singlets to $H_{2}^{\ast}$. This is similar
to having a single superfield pair $H_{1}$,$\bar{H}_{1}$ in a supersymmetric
model and is {\it not} analogous to the physics of our two Higgs pair vacua.
For this reason, we follow the second method; that is, we introduce a 
discrete symmetry that allows all quarks/leptons to couple to either 
$H_{1}$ or $H_{1}^{\ast}$, but forbids any Yukawa couplings of 
quarks/leptons to $H_{2}$ and $H_{2}^{\ast}$ at the classical level. Note that this discrete 
symmetry is the field theory analogue of the ``stringy'' $(p,q)$ and 
$[s,t]$ Leray selection rules for cubic Yukawa couplings
in our two Higgs pair vacua.

There are several discrete symmetries that can be imposed on our toy model to
implement the ``decoupling'' of $H_{2}$ from quark/leptons. The simplest of
these is a ${\mathbb Z}_{2}$ symmetry defined as follows.
Constrain the Lagranian to be invariant under 
the action
\begin{equation}
(\bar{Q},\bar{L})\longrightarrow (\bar{Q},\bar{L}), 
\quad (u,d,\nu,e) \longrightarrow (u,d,\nu,e)
\label{volker6}
\end{equation}
and
\begin{equation}   
\quad H_{1}\longrightarrow H_{1}, 
\quad H_{2} \longrightarrow -H_{2}, \quad \phi\longrightarrow -\phi.
\label{volker6A}
\end{equation}
Then, up to operators of dimension 4 in the fields, the Lagrangian  is restricted to be of the form 
\begin{equation}
  {\cal{L}}= 
  {\cal{L}}_\text{kinetic}+
  {\cal{L}}_\text{Yukawa}+
  {\cal{L}}_\text{potential},
\label{volker7}
\end{equation}
where ${\cal{L}}_{kinetic}$ is the canonically normalized gauged 
kinetic energy for all of the fields,
\begin{equation}
{\cal{L}}_\text{Yukawa}= {\lambda}^{1}_{u,ij}\bar{Q}_{i}{H}_{1}^{\ast}u_{j}+
        {\lambda}^{1}_{d,ij}\bar{Q}_{i}{H}_{1}d_{j}+
        {\lambda}^{1}_{\nu,ij}\bar{L}_{i}{H}_{1}^{\ast}\nu_{j}+
        {\lambda}^{1}_{e,ij}\bar{L}_{i}{H}_{1}e_{j} +hc
\label{volker8}
\end{equation}
with $i,j=1,2,3$ unrestrained and ${\cal{L}}_{potential}=-V$ with
\begin{equation}
V=V_{F}+V_{D}+{\cal{V}}
\label{ok1}
\end{equation}
such that 
\begin{equation}
V_{F}=\lambda_{1} ({H}_{1}^{\ast}\cdot H_{2})({H}_{2}^{\ast}\cdot H_{1}) +\lambda_{2} \big(({H}_{1}^{\ast}\cdot H_{2}) ({H}_{1}^{\ast}\cdot H_{2}) + ({H}_{2}^{\ast}\cdot H_{1}) ({H}_{2}^{\ast}\cdot H_{1})\big)
\label{ok2}
\end{equation}
\begin{equation}
V_{D}=\lambda_{3} \lvert H_{1}\rvert^{4}+\lambda_{4} \lvert H_{2}\rvert^{4}  +\lambda_{5} \lvert H_{1}\rvert^{2} \lvert H_{2}\rvert^{2}
\label{ok3}
\end{equation}
and
\begin{equation}
{\cal{V}}=-\mu_{1}^{2}\lvert H_{1}\rvert^{2} - \mu_{2}^{2}\lvert H_{2}\rvert^{2} 
    -\frac{{\mu}_{\phi}^{2}}{2} \phi^{2}+\rho_{3}\phi\big({H}_{1}^{\ast}\cdot H_{2}+ 
    {H}_{2}^{\ast}\cdot H_{1}\big) +\phi^{2} \big(\gamma_{1}\lvert H_{1}\rvert^{2} +\gamma_{2}
    \lvert H_{2}\rvert^{2} \big) + \rho_{4} \phi^{4} \; .
\label{volker9}
\end{equation}
Note that we have, for simplicity, taken $\lambda_{2}$ and $\rho_{3}$ to be real. For $V$ to be hermitian, all other coefficients in~\eqref{ok2},~\eqref{ok3} and~\eqref{volker9} must be real. Finally, to ensure vacuum stability we choose all coupling parameters to be positive.

In addition to the Yukawa couplings to $H_{2}$ being disallowed, the potentials $V_{F}$ and 
$V_{D} $ are also consistent with the potential energy of our two Higgs pair string vacuum. Specifically, the $F$-term contribution to the potential generated from the classical superpotential $W_{\mu}$ in~\eqref{burt3}, disregarding the terms with $\phi_{\bar{m}}$ and setting ${\bar{H}}_{1}$,
${\bar{H}}_{2}$ to be $H^{\ast}_{1}$, $H^{\ast}_{2}$ respectively for the reasons discussed previously, contains precisely the same terms as in $V_{F}$. They differ only in that their coefficients are related in the supersymmetric case, whereas $\lambda_{1}$, $\lambda_{2}$ in $V_{F}$ can be completely independent. Similarly, the $D$-term contribution to the supersymmetric potential, again setting 
${\bar{H}}_{1}$, ${\bar{H}}_{2}$ to be $H^{\ast}_{1}$, $H^{\ast}_{2}$, contains the same terms as in $V_{D}$, albeit with constrained coefficients. The coefficients $\lambda_{3}$, 
$\lambda_{4}$,  $\lambda_{5}$ in $V_{D}$ can be independent.

There are several other important, but more subtle, features of our two Higgs pair string
vacua that are captured in the remaining term $\cal{V}$ of the potential. First, recall that in
these string vacua quadratic mass terms do not appear for the Higgs fields 
since they are zero modes of the Dirac operator. However, supersymmetry
breaking and radiative corrections are expected to induce non-vanishing vacuum expectation 
values  for these fields. This symmetry breaking is modeled in our ${\mathbb{Z}}_{2}$ toy theory 
by the appearance of such mass terms in $\cal{V}$ with negative sign. To be consistent with electroweak breaking, we will choose parameters $\mu_{1}$, $\mu_{2}$ and $\lambda_{i}$, $i=1,\dots,5$ so that
\begin{equation}
\langle H_{1} \rangle \sim  \langle H_{2} \rangle \approx M_{EW} 
\label{not1}
\end{equation}
Second, moduli fields must have a vanishing perturbative
potential in string theory. However, non-perturbative effects and supersymmetry
breaking are expected to induce a moduli potential leading to stable,
non-zero moduli expectation values. This is modeled in our toy theory
by the the pure $\phi^{2}$ and $\phi^{4}$ terms in $\cal{V}$. Since
$\phi$ represents moduli with potentially large expectation values, we will
choose parameters $\mu_{\phi}$ and $\rho_{4}$ so that 
\begin{equation}
\langle\phi\rangle \lesssim M_{c} \; .
\label{not2}
\end{equation}
Finally, note that the ${\mathbb{Z}}_{2}$ symmetry allows mixed cubic and quartic $\phi$-$H$ couplings in ${\cal{V}}$. Such cubic terms cannot arise from a cubic superpotential. Quartic
terms might occur, but are disallowed by the $(p,q)$ and $[s,t]$ selection 
rules of our string vacua. However, both terms can be expected to arise in the string potential energy after supersymmetry breaking, radiative corrections and non-perturbative effects are taken into account.
To ensure that these terms are consistent with electroweak symmetry breaking~\eqref{not1} and the 
large modulus expectation value~\eqref{not2}, one must choose
coefficients $\rho_{3}$ and $\gamma_{1}$,$\gamma_{2}$ to satisfy
\begin{equation}
\rho_{3} \sim \big(\frac{M_{EW}}{M_{c}}\big)M_{EW} \; , \quad
\gamma_{1},\gamma_{2} \sim  \big(\frac{M_{EW}}{M_{c}}\big)^{2} \; .
\label{mike6}
\end{equation}
From the point of view of the toy model with ${\mathbb Z}_{2}$ discrete symmetry, this is
fine-tuning of the coefficients. However, it is a natural requirement if we want our toy model
to reflect the appropriate electroweak symmetry breaking in the two Higgs pair string vacua.

Of course, there is an infinite set of operators that are of order dimension five and higher in the fields that are consistent with the ${\mathbb{Z}}_{2}$ discrete symmetry. Here, we will be interested only in the dimension five operators 
\begin{equation}
{\cal{L}}_{5}= {\tilde{\lambda}}_{u,ij}\frac{\phi}{M_{c}}\bar{Q}_{i}{H}_{2}^{\ast}u_{j}+
        {\tilde{\lambda}}_{d,ij}\frac{\phi}{M_{c}}\bar{Q}_{i}{H}_{2}d_{j}+
        {\tilde{\lambda}}_{\nu,ij}\frac{\phi}{M_{c}}\bar{L}_{i}{H}_{2}^{\ast}\nu_{j}+
        {\tilde{\lambda}}_{e,ij}\frac{\phi}{M_{c}}\bar{L}_{i}{H}_{2}e_{j}+hc
\label{aa1}
\end{equation}
related to flavor-changing neutral currents. Note that a non-vanishing vacuum expectation value $\langle  \phi \rangle \neq 0$ will induce Yukawa couplings of the quarks/leptons to the the second Higgs doublet $H_{2}$ of the form
\begin{equation}
{\cal{L}}_{5, Yukawa}= {\lambda^{2}}_{u,ij}\bar{Q}_{i}{H}_{2}^{\ast}u_{j}+
      {\lambda^{2}}_{d,ij}\bar{Q}_{i}{H}_{2}d_{j}+
        {\lambda^{2}}_{\nu,ij}\bar{L}_{i}{H}_{2}^{\ast}\nu_{j}+
        {\lambda^{2}}_{e,ij}\bar{L}_{i}{H}_{2}e_{j} +hc \; ,
\label{aa2}
\end{equation}
where
\begin{equation}
{\lambda^{2}}_{u(\nu),ij}= {\tilde{\lambda}}_{u(\nu),ij}\frac{\langle \phi \rangle}{M_{c}} \; , \quad
{\lambda^{2}}_{d(e),ij}= {\tilde{\lambda}}_{d(e),ij}\frac{\langle \phi \rangle}{M_{c}} \; .
\label{aa3}
\end{equation}
Since one expects $\frac{\langle \phi \rangle}{M_{c}} < 1$, the Yukawa couplings to the second Higgs $H_{2}$ are naturally smaller that the couplings to $H_{1}$.
To be consistent with the two Higgs pair string vacua, it follows from~\eqref{home2} that we should choose
\begin{equation}
\lambda^{2}_{u(\nu),ij}  \ll \lambda^{1}_{u(\nu),ij} \;, \quad \lambda^{2}_{d(e),ij} \ll \lambda^{1}_{d(e),ij} \; . 
\label{aa4}
\end{equation}
More specifically, from~\eqref{burt8a},~\eqref{home4} and the associated discussion one might expect
\begin{equation}
10^{-7} \lambda^{1}_{u(\nu),ij}  \lesssim \lambda^{2}_{u(\nu),ij}  \lesssim 10^{-7} \;, 
\quad  10^{-7}  \lambda^{1}_{d(e),ij} \lesssim \lambda^{2}_{d(e),ij} \lesssim 10^{-7} \; .  
\label{aa5}
\end{equation}

\subsection{The Vacuum State}

To find the vacuum of this theory, one has to find the local minima of the 
potential $V$. To do this, define the 
component fields of the two Higgs doublets by
\begin{equation}
  H_{1}=\frac{1}{\sqrt{2}}\left(\begin{array}{c} h_{1} + i h_{2} \\ h_{3} + i h_{4} 
  \end{array}\right), \quad
  H_{2}=\frac{1}{\sqrt{2}}\left(\begin{array}{c} h_{5} + i h_{6} \\ h_{7} + i h_{8} 
  \end{array}\right)
\label{volker10}
\end{equation}
It turns out that for a generic choice of coefficients
there are several local minima. For simplicity of the analysis, we choose
the one most closely related to the standard model vacuum. The analytic expressions for the vacuum expectation values, as well as the scalar mass eigenvalues and eigenstates, greatly simplify if we take all coefficients $\lambda_{i}$, $i=1,\dots,5$ to have the  identical value $\lambda$. With this simplification, this local minimum is specified by
\begin{equation}
  \langle h_{3} \rangle = \frac{\mu_{1}}{\sqrt{\lambda}} \; , \quad  \langle h_{8} \rangle = \frac{\mu_{2}}{\sqrt{\lambda}} \;  , \quad \langle\phi\rangle = \frac{\mu_{\phi}}{2\sqrt{\rho_{4}}} 
 \label{ny1} 
\end{equation}
with all other expectation values vanishing. This vacuum clearly spontaneoously breaks $SU(3)_{C}\times SU(2)_{L}\times U(1)_{Y} \longrightarrow U(1)_{EM}$. Note that both Higgs doublets contribute to the mass matrix of the vector bosons. Despite this, as mentioned above, the GIM mechanism continues to apply  at tree level and all $Z$ mediated flavor-changing currents vanish.

The scalar mass matrix is easily evaluated and diagonalized in this vacuum. Expanding around the vacuum expectation values in~\eqref{ny1} and writing $h_{3}=\langle h_{3} \rangle +\bar{h}_{3}$, 
$h_{8}=\langle h_{8} \rangle +\bar{h}_{8}$ and $\phi=\langle \phi \rangle +{\bar{\phi}}$, we find that the square of the mass eigenvalues and the associated eigenstates are given respectively by 
\begin{equation}
\begin{array}{ll}
M^{2}_{h_{1}^{'}}=0,& 
M^{2}_{h_{2}^{'}}=0, \\
M^{2}_{h_{3}^{'}}=4 \mu_{1}^{2}, &
M^{2}_{h_{4}^{'}}=0,  \\
M^{2}_{h_{5}^{'}}= 4 ( \mu_{1}^{2}+\mu_{2}^{2}),  &
M^{2}_{h_{6}^{'}}=  \mu_{1}^{2}+\mu_{2}^{2}, \\
M^{2}_{h_{7}^{'}}= \mu_{1}^{2}+\mu_{2}^{2},  &
M^{2}_{h_{8}^{'}}= 4 \mu_{2}^{2}, \\
M_{\phi'}^{2}=2 \mu_{\phi}^{2} &
\label{prin1}
\end{array}
\end{equation}
and 
\begin{equation}
\begin{array}{ll}
h_{1}^{'}=-\tilde{\mu_{1}} h_{4}+\tilde{\mu_{2}}h_{7}, &
h_{2}^{'}=\tilde{\mu_{1}}h_{1}-\tilde{\mu_{2}}h_{6}, \\
h_{3}^{'}=\bar{h}_{3}, & 
h_{4}^{'}=\tilde{\mu_{1}}h_{2}+\tilde{\mu_{2}}h_{5}, \\
h_{5}^{'}= \tilde{\mu_{2}}h_{4}+\tilde{\mu_{1}}h_{7}, &
h_{6}^{'}=-\tilde{\mu_{2}}h_{1}-\tilde{\mu_{1}}h_{6}, \\
h_{7}^{'}=-\tilde{\mu_{2}}h_{2}+\tilde{\mu_{1}}h_{5}, &
h_{8}^{'}=\bar{h}_{8}, \\
\phi^{'}={\bar{\phi}} \; 
\label{prin2}
\end{array}
\end{equation}
where 
\begin{equation}
\tilde{\mu_{i}}=\frac{\mu_{i}}{\sqrt{\mu_{1}^{2}+\mu_{2}^{2}}}  \quad i=1,2 \; .
\label{kind1}
\end{equation}
Clearly $h^{'}_{1}$, $h^{'}_{2}$ and $h^{'}_{4}$, which can be rotated into the charged eigenstates 
\begin{equation}
{\cal{G}}^{0}=h_{1}^{'} \; , \quad {\cal{G}}^{\pm}=\frac{1}{\sqrt{2}}(h_{2}^{'} \pm \imath h_{4}^{'}) \; , 
 \label{book1}
 \end{equation}
are the Goldstone bosons. Since in the unitary gauge they will be absorbed into the longitudinal components of the $Z$  and $W^{\pm}$ vector bosons, we will
henceforth ignore these fields. The remaining Higgs scalars we group into charge eigenstates as
\begin{equation}
{\cal{H}}^{0}_{1}=h_{3}^{'} \; , \quad {\cal{H}}^{0}_{2}=h_{5}^{'} \; , \quad {\cal{H}}^{0}_{3}=h_{8}^{'} 
\label{prin3}
\end{equation}
and
\begin{equation}
{\cal{H}}^{\pm}=\frac{1}{\sqrt{2}}(h_{6}^{'}\pm \imath h_{7}^{'})  \; ,
\label{prin4}
\end{equation}
with masses 
\begin{equation}
M^{2}_{{\cal{H}}^{0}_{1}}=4 \mu_{1}^{2} \; , \quad M^{2}_{{\cal{H}}^{0}_{2}}= 4 ( \mu_{1}^{2}+\mu_{2}^{2}) \; , \quad M^{2}_{{\cal{H}}^{0}_{3}}= 4 \mu_{2}^{2}
\label{book2}
\end{equation}
and
\begin{equation}
M^{2}_{{\cal{H}}^{\pm}}= \mu_{1}^{2}+\mu_{2}^{2}
\label{book4}
\end{equation}
respectively. Since we are interested in flavor-changing neutral currents, we will ignore ${\cal{H}}^{\pm}$ and consider the currents coupling to ${\cal{H}}^{0}_{1}$, ${\cal{H}}^{0}_{2}$ and ${\cal{H}}^{0}_{3}$ only. The charge neutral field $\phi^{'}$ does mediate a flavor-changing neutral current. 
However, it will naturally be suppressed by the factor $\frac{\langle H_{2} \rangle}{M_{c}}$ and, hence, is 
negligible.

\subsection{Flavor-Changing Neutral Currents}

Having determined the vacuum state, we can expand the two Yukawa terms given in~\eqref{volker8} 
and~\eqref{aa2} to find the fermion mass matrices and the Higgs induced flavor-changing neutral interactions. For simplicity, we will always assume $\lambda^{1,2}_{u(\nu),ij}$ and 
$\lambda^{1,2}_{d(e),ij}$ are real and symmetric. First, consider the fermion mass matrices. For up-quarks, one finds
\begin{equation}
\big( {\cal{L}}_{Yukawa}+{\cal{L}}_{5,Yukawa} \big) |_{\rm up-mass}=
{\bar{U}_{i}} \big(\frac{\lambda^{1}_{u,ij}}{\sqrt{2}}\langle h_{3} \rangle -i\frac{\lambda^{2}_{u,ij}}{\sqrt{2}}\langle h_{8} \rangle \big)u_{j}+hc \; .
\label{by1}
\end{equation}
This can always be written in terms of a diagonal mass matrix and its eigenstates. For example, the first term becomes
\begin{equation}
{\bar{U}_{i}} \big(\frac{\lambda^{1}_{u,ij}}{\sqrt{2}}\langle h_{3} \rangle -i\frac{\lambda^{2}_{u,ij}}{\sqrt{2}}\langle h_{8} \rangle \big)u_{j}=
{\bar{\tilde{U}}}_{i}{\cal{M}}^{\rm diag}_{u,ij}{\tilde{u}}_{j} \; ,
\label{by2}
\end{equation}
which allows us to re-express
\begin{equation}
\frac{\lambda^{1}_{u,ij}}{\sqrt{2}}{\bar{U}_{i}} u_{j}=
{\bar{\tilde{U}}}_{i}\frac{{\cal{M}}^{\rm diag}_{u,ij}}{\langle h_{3} \rangle}{\tilde{u}}_{j} +i\frac{\lambda^{2}_{u,ij}}{\sqrt{2}}\frac{\langle h_{8} \rangle}{\langle h_{3} \rangle} 
{\bar{\tilde{U}}_{i}} {\tilde{u}}_{j} \; .
\label{by3}
\end{equation}
Note that, in the last term, we have replaced ${\bar{U}_{i}} ,u_{j}$ by the eigenstates 
${\bar{\tilde{U}}_{i}} , {\tilde{u}}_{j}$. This is valid to leading order since it follows from~\eqref{not1} and~\eqref{aa4} that 
\begin{equation}
\lambda^{2}_{u,ij}\langle h_{8} \rangle \ll \lambda^{1}_{u,ij} \langle h_{3} \rangle \; .
\label{by4}
\end{equation}
Similar expressions hold for the hermitian conjugate terms, down-quarks and the $\nu , e$-leptons.

One can now evaluate the flavor-changing neutral interactions. For up-quarks, we find that
\begin{equation}
\big( {\cal{L}}_{Yukawa}+{\cal{L}}_{5,Yukawa} \big) |_{\rm up-neutral}=
\frac{\lambda^{2}_{u,ij}}{\sqrt{2}} {\bar{\tilde{U}}_{i}} \big( i\frac{\langle h_{8} \rangle}{\langle h_{3} \rangle} ({\bar{h}}_{3}-ih_{4})+(h_{7}-i{\bar{h}}_{8}) \big) {\tilde{u}}_{j} +hc \; ,
\label{by5}
\end{equation}
where we have used expression~\eqref{by3} and dropped the flavor-diagonal ${\cal{M}}^{\rm diag}_{u,ij}$ term.
From~\eqref{ny1},~\eqref{prin2} and~\eqref{prin3}, one can write~\eqref{by5} in terms of the neutral Higgs eigenstates. The result is
\begin{equation}
\big( {\cal{L}}_{Yukawa}+{\cal{L}}_{5,Yukawa} \big) |_{\rm up-neutral}=
\frac{\lambda^{2}_{u,ij}}{\sqrt{2}} {\bar{\tilde{U}}_{i}} \big( i\frac{{\tilde{\mu}}_{2}}{{\tilde{\mu}}_{1}} {\cal{H}}^{0}_{1}+ \frac{1}{{\tilde{\mu}}_{1}}{\cal{H}}^{0}_{2}-i{\cal{H}}^{0}_{3} \big) {\tilde{u}}_{j} +hc \; .
\label{by6}
\end{equation}
Written in terms of the Dirac spinors
\begin{equation}
q_{u,i}={\tilde{U}}_{i} \oplus {\tilde{u}}_{i} \; ,
\label{by7}
\end{equation}
this becomes
\begin{equation}
\big( {\cal{L}}_{Yukawa}+{\cal{L}}_{5,Yukawa} \big) |_{\rm up-neutral}=
\frac{\lambda^{2}_{u,ij}}{\sqrt{2}}  \big( -i\frac{{\tilde{\mu}}_{2}}{{\tilde{\mu}}_{1}} ({\bar{q}}_{u,i}\gamma^{5}q_{u,j}) {\cal{H}}^{0}_{1}+ \frac{1}{{\tilde{\mu}}_{1}} ({\bar{q}}_{u,i}q_{u,j}) {\cal{H}}^{0}_{2}+i ({\bar{q}}_{u,i}\gamma^{5}q_{u,j}){\cal{H}}^{0}_{3} \big) \; .
\label{by8}
\end{equation}
Similar expressions hold for the down-quarks and $\nu,e$-leptons. Putting everything together, we find that the flavor-changing neutral interactions are given by
\begin{equation}
\big( {\cal{L}}_{Yukawa}+{\cal{L}}_{5,Yukawa} \big) |_{\rm neutral}=
{\cal{J}}^{1}{\cal{H}}^{0}_{1}+{\cal{J}}^{2}{\cal{H}}^{0}_{2}+{\cal{J}}^{3}{\cal{H}}^{0}_{3} \; ,
\label{by9}
\end{equation}
where
\begin{equation}
{\cal{J}}^{1}=-i\frac{\lambda^{2}_{u(\nu),ij}}{\sqrt{2}}\frac{{\tilde{\mu}}_{2}}{{\tilde{\mu}}_{1}} ({\bar{q}}_{u(\nu),i}\gamma^{5}q_{u(\nu),j}) +i\frac{\lambda^{2}_{d(e),ij}}{\sqrt{2}}\frac{{\tilde{\mu}}_{2}}{{\tilde{\mu}}_{1}} ({\bar{q}}_{d(e),i}\gamma^{5}q_{d(e),j}) \; , 
\label{by10}
\end{equation}
\begin{equation}
{\cal{J}}^{2}=\frac{\lambda^{2}_{u(\nu),ij}}{\sqrt{2}}\frac{1}{{\tilde{\mu}}_{1}} ({\bar{q}}_{u(\nu),i}q_{u(\nu),j}) +\frac{\lambda^{2}_{d(e),ij}}{\sqrt{2}}\frac{1}{{\tilde{\mu}}_{1}} ({\bar{q}}_{d(e),i}q_{d(e),j})  \; ,
\label{by11}
\end{equation}
\begin{equation}
{\cal{J}}^{3}=i\frac{\lambda^{2}_{u(\nu),ij}}{\sqrt{2}} ({\bar{q}}_{u(\nu),i}\gamma^{5}q_{u(\nu),j}) -i\frac{\lambda^{2}_{d(e),ij}}{\sqrt{2}} ({\bar{q}}_{d(e),i}\gamma^{5}q_{d(e),j}) \; . 
\label{by11}
\end{equation}
Note that these flavor-changing currents all vanish as $\lambda^{2}_{u(\nu),ij}, \lambda^{2}_{d(e),ij} \rightarrow 0$, as they must.



\subsection{Phenomenology}

The most stringent bounds on Higgs mediated flavor changing neutral currents arise from 
the experimental data on the mass splitting of neutral pseudoscalar $F^{0}-{\bar{F}}^{0}$ meson
eigenstates. Theoretically, the mass difference $\Delta M_{F}$ is given by
\begin{equation}
M_{F}\Delta M_{F}= |\langle F^{0}| {\cal{L}}_{\rm eff}|{\bar{F}}^{0}  \rangle | \; ,
\label{one1}
\end{equation}
where ${\cal{L}}_{\rm eff}$ is the low energy $\Delta F=2$ effective Lagrangian arising from a variety of processes~\cite{atwood1,atwood2}. First, there is a well-known contribution from the standard model part of our  simplified theory. In addition, we have terms rising from the flavor-changing neutral Higgs vertices in~\eqref{by9}-\eqref{by11}. These lead to the tree-level graphs shown in Figure 3 which, at low energy, give extra contributions to the mass splitting. Using the results of~\cite{atwood1}, we find that the Higgs mediated flavor changing neutral currents lead to an additional contribution to the mass splitting given by

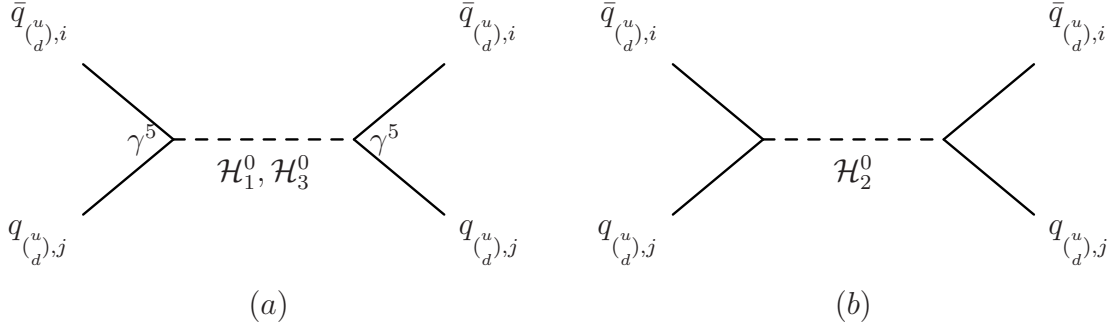
\begin{figure}
  \centering
\begin{fmffile}{treelevel}
  \newenvironment{treegraph2}
  {\begin{fmfgraph*}(60,20)
    \fmfleft{i1,i2}
    \fmfright{o1,o2}
    \fmf{vanilla}{i1,v1,i2}
    \fmf{vanilla}{o1,v2,o2} }
  {\end{fmfgraph*}}
 $\begin{array}{c c c}
\\
\\
\fmfframe(0,0)(0,0){
  \begin{treegraph2}
    \fmflabel{$\bar{q}_{(\tiny{\begin{array}{@{}c@{}}u \\ d\end{array}}),i}$}{i2}
    \fmflabel{$q_{(\tiny{\begin{array}{@{}c@{}}u \\ d\end{array}}),j}$}{i1}
    \fmflabel{$\bar{q}_{(\tiny{\begin{array}{@{}c@{}}u \\ d\end{array}}),i}$}{o2}
    \fmflabel{$q_{(\tiny{\begin{array}{@{}c@{}}u \\ d\end{array}}),j}$}{o1}
    \fmflabel{$\gamma^{5}$}{v1}
    \fmflabel{$\gamma^{5}$}{v2}
    \fmf{dashes,label=${\cal{H}}^{0}_{1},,{\cal{H}}^{0}_{3}$}{v1,v2}
  \end{treegraph2}
}
&
 \hspace{10mm}
&
  \begin{treegraph2}
    \fmflabel{$\bar{q}_{(\tiny{\begin{array}{@{}c@{}}u \\ d\end{array}}),i}$}{i2}
    \fmflabel{$q_{(\tiny{\begin{array}{@{}c@{}}u \\ d\end{array}}),j}$}{i1}
    \fmflabel{$\bar{q}_{(\tiny{\begin{array}{@{}c@{}}u \\ d\end{array}}),i}$}{o2}
    \fmflabel{$q_{(\tiny{\begin{array}{@{}c@{}}u \\ d\end{array}}),j}$}{o1}

    \fmf{dashes,label=${\cal{H}}^{0}_{2}$}{v1,v2}
  \end{treegraph2}
\\  \\ (a) & & (b)
 \end{array} $
\end{fmffile}
   \caption{Feynman diagrams of the tree level contributions to neutral
meson mixing mediated by Higgs bosons. Note that graphs (a) and (b)
involve pseudoscalar and scalar interactions respectively.}
  \label{fig:treelevel}
\end{figure}

\begin{equation}
M_{F}\Delta M_{F}^{FCNC}= \frac{B_{F}}{8} \big({\lambda^{2}}_{(u,d),ij} \big)^{2} 
\big[ (\pm) \{ \big( \frac{\mu_{2}}{\mu_{1}} \big)^{2} \frac{1}{\mu_{1}^{2}} 
- \frac{1}{\mu_{2}^{2}} \} {\cal{P}}^{F}_{ij} + \frac{1}{\mu_{1}^{2}} {\cal{S}}^{F}_{ij}
\big] \; ,
\label{one2}
\end{equation}
where 
\begin{equation}
{\cal{P}}^{F}_{ij}=-\frac{f_{F}^{2}M_{F}^{2}}{6}\big(1+\frac{11M_{F}^{2}}{(m_{i}+m_{j})^{2}}\big) \; , \quad
{\cal{S}}^{F}_{ij} =\frac{f_{F}^{2}M_{F}^{2}}{6}\big(1+\frac{M_{F}^{2}}{(m_{i}+m_{j})^{2}}\big) 
\label{one3}
\end{equation}
are associated with the pseudoscalar and scalar interaction graphs, Figure 3(a) and Figure 3(b), respectively. Here $f_{F}$ is the pseudoscalar decay constant, $M_{F}$ is the leading order meson mass, $m_{i}$ is the mass of the $i$-th constituent quark and $B_{F}$ is the $B$-parameter of the vacuum insertion approximation defined in~\cite{atwood1}. The label $(u,d)$ tells one to choose the $\lambda$ coefficient associated with the up-quark or down-quark content of the meson $F$ and the indices $i,j$, where $i \neq j$,  indicate which two families compose $F$. In this paper, we simplify the analysis by considering two natural limits of~\eqref{one2}, each consistent with all previous assumptions. The first limit is to take $\mu_{2}=\mu_{1} \approx M_{EW}$. Expression~\eqref{one2} then simplifies to 
\begin{equation}
M_{F}\Delta M_{F}^{FCNC(I)}= \frac{B_{F}}{8} \big({\lambda^{2}}_{(u,d),ij} \big)^{2} 
\frac{1}{M_{EW}^{2}} {\cal{S}}^{F}_{ij} \; .
\label{one4}
\end{equation}
As a second limit, let us assume that $\mu_{2} \ll \mu_{1} \approx M_{EW}$. In this case, 
the $\mu_{1}$ contribution is sub-dominant and~\eqref{one2} becomes
\begin{equation}
M_{F}\Delta M_{F}^{FCNC(II)}= \mp  \frac{B_{F}}{8} \big({\lambda^{2}}_{(u,d),ij} \big)^{2} 
\frac{1}{\mu_{2}^{2}} {\cal{P}}^{F}_{ij} \; ,
\label{one5}
\end{equation}
which can be written as
\begin{equation}
\Delta M_{F}^{FCNC(II)}= \mp \Delta M_{F}^{FCNC(I)} \big(\frac{M_{EW}^{2}}{\mu_{2}^{2}}\big) \big(\frac{{\cal{P}}^{F}_{ij}}{{\cal{S}}^{F}_{ij}}\big) \; .
\label{one6}
\end{equation}
It follows from~\eqref{one3} that, in general, $\frac{ |{\cal{P}}^{F}_{ij}|}{{\cal{S}}^{F}_{ij}} \sim 10$
and from our assumption that $\frac{M_{EW}^{2}}{\mu_{2}^{2}} \gg 1$. Hence,
\begin{equation}
|\Delta M_{F}^{FCNC(II)}| \gg \Delta M_{F}^{FCNC(I)} \; .
\label{one7}
\end{equation}
We will analyze the implications of both limits. Before proceeding, recall from~\eqref{aa5} that a natural range for the the Yukawa coefficients $ \lambda^{2}_{(u,d),ij}$ is
\begin{equation}
10^{-7} \lambda^{1}_{(u,d),ij}  \lesssim \lambda^{2}_{(u,d),ij}  \lesssim 10^{-7} \; .
\label{one8}
\end{equation}
There are various ways to estimate the flavor non-diagonal coefficients 
$\lambda^{1}_{(u,d),ij} $, $i \neq j$. Here, we will simply assume each is of the same order of magnitude as the largest diagonal Yukawa coupling of the $u$ or $d$ type corresponding to the $i$ and $j$ families. Other commonly used estimates simply strengthen our conclusions.

\begin{table}
 \centering
 \begin{tabular*}{\textwidth}{@{\extracolsep{\fill}}|c|c|c|c||c|c|}
  \hline
  $F^{0}$ & ${\cal{P}}^{F}$ & ${\cal{S}}^{F}$ & $B_{F}$ & $\Delta M_{F}^{SM}$ & $\Delta M_{F}^{Exp}$ \\
  \hline
  $K^{0}$ & -27.5 & 2.5 & 0.75 & $1.4-4.6\times 10^{-15}$ & $3.51\times 10^{-15}$ \\
  $B_{d}^{0}$ & -2.65 & 0.37 & 1 & $10^{-13}-10^{-12}$ & $3.26\times 10^{-13}$ \\
  $D^{0}$ & -0.52 & 0.068 & 1 & $10^{-17}-10^{-16}$ & $ < 1.32\times 10^{-13}$ \\
  \hline

\end{tabular*}

\caption{Table of data pertinent to the calculation of $\Delta M_{F}$.  The data in the first
two columns have dimensions $GeV^{4}$, those in column three are dimensionless while the entries in the last two
columns are in $GeV$.}
\end{table}

In this paper, we will consider the $F^{0}$ mesons $K^{0}={\bar{s}}d$, $B_{d}^{0}={\bar{b}}d$ and $D^{0}={\bar{c}}u$, since their mass mixings with their conjugates are the best measured. The values for
${\cal{P}}^{F}_{ij}$, ${\cal{S}}^{F}_{ij}$ and $B_{F}$ for each of these mesons are presented in Table 1.
In addition, the last two columns of Table 1 contain the theoretical standard model contribution and the experimental value of $\Delta M_{F}$ respectively.
First consider $K^{0}-{\bar{K}}^{0}$ mixing. In the limit that $\mu_{2}=\mu_{1}\approx M_{EW}$, it follows from~\eqref{one4}, Table 1 and $M_{K^{0}}=.497GeV$ that
\begin{equation}
\Delta M_{K}^{FCNC(I)} \approx 4.72 \times 10^{-5} (\lambda^{2}_{d,12})^{2} GeV \; .
\label{one9}
\end{equation}
Assuming that $\lambda^{1}_{d,12} \sim \lambda^{1}_{s} \sim 10^{-4}$, the range~\eqref{one8} becomes
\begin{equation}
10^{-11} \lesssim \lambda^{2}_{d,12} \lesssim 10^{-7}
\label{one10}
\end{equation}
and, hence, 
\begin{equation}
4.72 \times 10^{-27} GeV \lesssim \Delta M_{K}^{FCNC(I)} \lesssim 4.72 \times 10^{-19} GeV \; .
\label{one11}
\end{equation}
This sits comfortably below the upper bound 
\begin{equation}
\Delta M_{K}^{FCNC} \lesssim 10^{-15} GeV
\label{one12}
\end{equation}
obtained using the $K^{0}$ entries in the last two columns of Table 1. Next, consider the second limit where $\mu_{2} \ll \mu_{1} \approx M_{EW}$. In this case, we know from~\eqref{one7} that this choice of parameters will come closer to saturating the upper bound. Using~\eqref{one6} and Table 1 we find
that
\begin{equation}
|\Delta M_{K}^{FCNC(II)}|= \Delta M_{K}^{FCNC(I)} \big(\frac{1.1 \times 10^{5} GeV^{2}}{\mu_{2}^{2}}\big)
\label{one13}
\end{equation}
If, for example, we take 
\begin{equation}
\mu_{2} \approx 7 GeV \; ,
\label{one14}
\end{equation}
corresponding to an ${\cal{H}}^{0}_{3}$ mass of $14GeV$, then it follows from~\eqref{one11} and~\eqref{one13} that
\begin{equation}
10^{-23} GeV \lesssim  |\Delta M_{K}^{FCNC(II)}| \lesssim 10^{-15} GeV \; .
\label{one15}
\end{equation}
The choice of $\mu_{2}$ in~\eqref{one14} is purely illustrative, chosen so that the Higgs mediated flavor changing currents can induce $K^{0}$ mixing of the same order as the experimental data. A more detailed study of our theory would be required to determine if a neutral Higgs boson can be this light relative to the electroweak scale. Of course, if the mass of ${\cal{H}}^{0}_{3}$ is larger, its contribution to neutral meson mixing would rapidly decrease.
We conclude that if $\lambda^{2}_{d,12}$  saturates its upper bound of $10^{-7}$ and the  neutral Higgs ${\cal{H}}^{0}_{3}$ is sufficiently light, then the contribution of the Higgs mediated flavor-changng neutral currents can play a measurable role in $K^{0}-{\bar{K}}^{0}$ mixing.

Next, let us discuss $B_{d}^{0}-{\bar{B}}_{d}^{0}$ mixing. In the limit that $\mu_{2}=\mu_{1}\approx M_{EW}$, it follows from~\eqref{one4}, Table 1 and $M_{B_{d}^{0}}=5.28 GeV$ that
\begin{equation}
\Delta M_{B_{d}}^{FCNC(I)} \approx .876 \times 10^{-6} (\lambda^{2}_{d,13})^{2} GeV \; .
\label{one16}
\end{equation}
Assuming that $\lambda^{1}_{d,13} \sim \lambda^{1}_{b} \sim 10^{-2}$, the range~\eqref{one8} becomes
\begin{equation}
10^{-9} \lesssim \lambda^{2}_{d,13} \lesssim 10^{-7}
\label{one17}
\end{equation}
and, hence, 
\begin{equation}
.876 \times 10^{-24} GeV \lesssim \Delta M_{B_{d}}^{FCNC(I)} \lesssim .876 \times 10^{-20} GeV \; .
\label{one18}
\end{equation}
This contribution is well below the upper bound of
\begin{equation}
\Delta M_{B_{d}}^{FCNC} \lesssim 10^{-13}GeV
\label{one19}
\end{equation}
obtained using the $B_{d}^{0}$ entries in the last two columns of Table 1. Next, consider the second limit where $\mu_{2} \ll \mu_{1} \approx M_{EW}$. In this case, we know from~\eqref{one7} that this choice of parameters will come closer to saturating the upper bound. Using~\eqref{one6} and Table 1 we find
that
\begin{equation}
|\Delta M_{B_{d}}^{FCNC(II)}|= \Delta M_{B_{d}}^{FCNC(I)} \big(\frac{7.16 \times 10^{4} GeV^{2}}{\mu_{2}^{2}}\big)
\label{one20}
\end{equation}
If we take, for example, 
\begin{equation}
\mu_{2} \approx 7 GeV \; ,
\label{21}
\end{equation}
thus saturating the upper bound in the $K^{0}$ case, then it follows from~\eqref{one18} and~\eqref{one20} that
\begin{equation}
1.28 \times 10^{-21} GeV \lesssim  |\Delta M_{B_{d}}^{FCNC(II)}| \lesssim 1.28 \times 10^{-17} GeV \; .
\label{one22}
\end{equation}
We conclude that even if $\lambda^{2}_{d,13}$  saturates its upper bound of $10^{-7}$ and the  neutral Higgs ${\cal{H}}^{0}_{3}$ is sufficiently light to saturate the upper bound in the $K^{0}$ case, the contribution of the Higgs mediated flavor-changing neutral currents to $B_{d}^{0}-{\bar{B_{d}}}^{0}$ mixing remains well below the presently measured upper bound.

Finally, consider the $D^{0}-{\bar{D}}_{0}$ case. If we assume that $\lambda^{1}_{u,12} \sim \lambda^{1}_{c} \sim 5 \times10^{-3}$, the range~\eqref{one8} becomes
\begin{equation}
5 \times 10^{-10} \lesssim \lambda^{2}_{u,12} \lesssim 10^{-7} \; .
\label{one23}
\end{equation}
It follows from this,~\eqref{one4}, Table 1 and $M_{D^{0}}=1.86GeV$ that in the limit that $\mu_{1}=\mu_{2} \approx M_{EW}$
\begin{equation}
1.14 \times 10^{-25} GeV \lesssim \Delta M_{D}^{FCNC(I)} \lesssim 4.56 \times 10^{-21} GeV \; ,
\label{one24}
\end{equation}
well below the upper bound of
\begin{equation}
\Delta M_{D}^{FCNC} \lesssim 10^{-13}GeV
\label{one25}
\end{equation}
obtained using the $D^{0}$ entries in the last two columns of Table 1. Finally, consider the second limit where $\mu_{2} \ll \mu_{1} \approx M_{EW}$. In this case, using~\eqref{one6}, Table 1,~\eqref{one24}
and $\mu_{2} \approx 7GeV$, we obtain
\begin{equation}
1.77 \times 10^{-22} GeV \lesssim  |\Delta M_{D}^{FCNC(II)}| \lesssim 7.11 \times 10^{-18} GeV \; .
\label{one26}
\end{equation}
We conclude that even if $\lambda^{2}_{u,12}$  saturates its upper bound of $10^{-7}$ and the  neutral Higgs ${\cal{H}}^{0}_{3}$ is sufficiently light to saturate the upper bound in the $K^{0}$ case, the contribution of the Higgs mediated flavor-changing neutral currents to $D^{0}-{\bar{D}}^{0}$ mixing remains well below the presently measured upper bound.

\section{Acknowledgments}

This research was supported by the Department of Physics and the
Math/Physics Research Group at the University of Pennsylvania under cooperative
research agreement DE-FG02-95ER40893 with the U. S. Department of Energy, and
an NSF Focused Research Grant DMS0139799 for "The Geometry of Superstrings".


\end{document}